\pdfoutput=1

\documentclass[11pt]{article}

\usepackage[final]{acl}

\usepackage{times}
\usepackage{latexsym}

\usepackage[T1]{fontenc}

\usepackage[utf8]{inputenc}

\usepackage{microtype}

\usepackage{inconsolata}

\usepackage{graphicx}

\usepackage{booktabs}
\usepackage{makecell}
\usepackage{multirow}
\usepackage{amsmath}
\usepackage{xcolor}      
\usepackage{colortbl}    
\usepackage{subcaption}  
\usepackage{enumitem}

\newcommand{\eg}{\hbox{\emph{e.g.}}}
\newcommand{\ie}{\hbox{\emph{i.e.}}}

\definecolor{CommentGreen}{HTML}{008000}

\definecolor{bg}{HTML}{F8F9FB}
\definecolor{bgc}{HTML}{FCF6E4}
\usepackage[framemethod=TikZ]{mdframed}
\mdfdefinestyle{MyFrame}{%
    linecolor=black,
    outerlinewidth=0.3pt,
    roundcorner=5pt,
    skipabove = 0.8\baselineskip,
    skipbelow = 0pt,
    innertopmargin=5.5pt, 
    innerbottommargin=3.5pt,
    innerrightmargin=5.5pt,
    innerleftmargin=5.5pt,
    backgroundcolor=bg,
    nobreak=false,
}

\title{Across Programming Language Silos: A Study on Cross-Lingual Retrieval-Augmented Code Generation}
\author{
    Qiming Zhu\textsuperscript{\rm 1,\rm 2},
    Jialun Cao\textsuperscript{\rm 3}\thanks{Corresponding Authors.},
    Xuanang Chen\textsuperscript{\rm 1\textasteriskcentered},
    Weili Zhang\textsuperscript{\rm 1},
    Yaojie Lu\textsuperscript{\rm 1},\\
    \textbf{Hongyu Lin\textsuperscript{\rm 1},}
    \textbf{Xianpei Han\textsuperscript{\rm 1},}
    \textbf{Le Sun\textsuperscript{\rm 1},}
    \textbf{Shing-Chi Cheung}\textsuperscript{\rm 3}
\\
\normalsize
    \textsuperscript{\rm 1}Chinese Information Processing Laboratory, Institute of Software, Chinese Academy of Sciences\\
\normalsize
    \textsuperscript{\rm 2}University of Chinese Academy of Sciences\\
\normalsize
    \textsuperscript{\rm 3}The Hong Kong University of Science and Technology\\
\normalsize
    \texttt{\{zhuqiming2022,chenxuanang,weili,luyaojie,hongyu,xianpei,sunle\}@iscas.ac.cn} \\
\normalsize
    \texttt{\{jcaoap,scc\}@cse.ust.hk}
}

\begin{document}

\maketitle

\begin{abstract}

Current research on large language models (LLMs) with retrieval-augmented code generation (RACG) has largely focused on single-language settings, leaving their cross-lingual effectiveness underexplored. Multilingual RACG systems are increasingly important for migrating and reusing code across programming languages (\textit{PL}s), a common yet challenging task in modern software development. To systematically study cross-lingual code knowledge transfer in RACG, we construct a dataset covering 13 \textit{PL}s with nearly 14K instances. 
Our experiments reveal three key insights: (1) Knowledge transfer in RACG across \textit{PL}s is non-trivial even using direct injection. (2) RACG exhibits unequal cross-lingual knowledge transfer, and its efficacy depends on linguistic affinity of \textit{PL} pair and diversity of LLM pretraining corpus. (3) RACG shows limited reliance on natural language information embedded in code when equipped with a code-specific retriever. These findings provide practical guidance for designing effective multilingual RACG systems.~\url{https://github.com/icip-cas/Cross-Lingual-RACG}.

\end{abstract}
\section{Introduction}

With the emergence and rapid development of large language models (LLMs), significant progress has been made in natural language (NL) to code generation task~\cite{chen2021evaluating, roziere2023codellama}.
Despite these advances, code-oriented LLMs still struggle to generate reliable and correct programs in complex real-world scenarios, largely due to limitations in model capacity and reasoning ability~\cite{zhang2023repocoder, jimenez2024swebench}.
Retrieval-Augmented Generation (RAG)~\cite{lewis2020rag, guu2020ragmodel} has emerged as a powerful paradigm to enhance LLMs, and demonstrates efficacy in code task~\cite{tan2024prompt}. 
Building on this paradigm, Retrieval-Augmented Code Generation (RACG) extends RAG to code generation by retrieving relevant code as external knowledge.

In particular, advancing cross-lingual RACG is increasingly necessary in modern software development, where code reuse and migration across programming languages are common yet remain challenging.
A key obstacle lies in the asymmetric distribution of code knowledge across programming languages (\textit{PL}s)~\cite{yang2024multilangse,cao2025buildbenchmarkrevisiting274}. 
Widely adopted \textit{PL}s like Python benefit from extensive documentation, active communities, and abundant repositories, whereas less popular \textit{PL}s such as Scala often suffer from sparse resources and limited maintenance~\cite{djurdjev2024popularity}. This disparity creates inconvenience for developers working with less popular \textit{PL}s.
At the same time, the demand for multilingual RACG is further amplified as enterprises modernize their technology stacks~\cite{yang2024multilangse}. Migrating to emerging \textit{PL}s can offer performance optimization and security compliance, creating demand for cross-lingual code transformation tools~\cite{mayer2017multisurvey, cassano2024knowledge}.

Although existing works~\cite{wang2024coderagbench, gao2024preference, parvez2021retrieval} explore the RACG, they are confined to limited \textit{PL}s such as Python and Java, leaving RACG across multiple \textit{PL}s and cross-lingual code knowledge transfer unexplored.
In this paper, we study the cross-lingual knowledge transfer in RACG and aim to answer 3 research questions (RQs) about it. Our investigation shows knowledge transfer performance of current code LLMs, while uncovering interesting findings for cross-lingual RACG.

\textbf{RQ1. How effective is RACG with direct oracle cross-\textit{PL} code injection?}
We investigates the efficacy upper bound of cross-lingual knowledge transfer for code generation tasks under idealized conditions, implemented via RACG with oracle retrieval, which is one of the direct knowledge injection. To strictly control variables, we conducted preliminary injection experiments on a \textit{Small but Parallel Multi-lingual Code Dataset}, which contains only 5 \textit{PL}s and 164 questions per \textit{PL}. We find that while cross-lingual knowledge transfer is feasible by RACG, it is still a non-trivial process with fundamental limitations, motivating further exploration of the knowledge transfer mechanisms.

\textbf{RQ2. How effective is RACG at cross-\textit{PL} knowledge transfer via code retrieval?}
To investigate the effectiveness of cross-lingual knowledge transfer in RACG, we construct a \textit{Large Multilingual Code Dataset} covering 13 \textit{PL}s. Each instance includes a NL prompt, a verified reference solution, and executable test cases. Using a complete RACG pipeline, we systematically evaluate enhancement from a source \textit{PL} to a different target \textit{PL}. Our experiments reveal that RACG enables unequal knowledge transfer—its efficacy depends to some extent on the linguistic affinity between the PL pair and the diversity of the LLM’s pretraining corpus. Notably, while multi-lingual LLMs show the ability of knowledge transfer, python-specialized LLMs struggle to leverage cross-lingual knowledge in context, underscoring the importance of broad pretraining for effective knowledge transfer.

\textbf{RQ3. How does NL information in code affect cross-\textit{PL} knowledge transfer in RACG?}
In this RQ, we simulate a setting where the retrieval corpus consists of pure code snippets without NL descriptions—reflecting the common scenario of isolated code fragments online. Our experiments show that removing NL from the corpus has only a marginal effect on cross‑lingual knowledge transfer performance, with overall performance declining very slightly. We also systematically evaluate three retrieval paradigms and domain‑specific code retrievers outperform, which suggests that while the existence of NL is not essential for enabling knowledge transfer, specialized code retrieval is critical for effectively bridging NL intent and code semantics in RACG applications.

The contributions are summarized as follows:
\begin{itemize}[leftmargin=*]
    \item We present the first study to explore the code knowledge transfer in RACG across multiple \textit{PL}s, which sheds light on the possibility of going across \textit{PL} silos.
    \item Our study deepens the understanding of cross-lingual knowledge transfer in RACG via extensive experiments (3 retrieval experimental settings and 5 code LLMs and 13 \textit{PL}s).
    \item We construct nearly 14k high-quality code generation instances and document annotations spanning 13 \textit{PL}s. We also plan to release the dataset and experiment code to boost the transparency and facilitate further study.
\end{itemize}

\section{Study Design}

\subsection{Task Formulation}

We formulate the problem as follows:
Let $L$ denote the set of \textit{PL}s. For $\text{RACG}_{l_{\text{src}} \to l_{\text{tgt}}}$ where $l_{\text{src}}, l_{\text{tgt}} \in L$, which is a RACG task that transfers knowledge from the code documentation corpus in the source \textit{PL} $l_{\text{src}}$ to the target \textit{PL} $l_{\text{tgt}}$ for code generation, the process is formalized as:
\begin{equation}
G\Big(p(q, l_{\text{tgt}}),\ \underbrace{R(D_{l_{\text{src}}}, q)}_{\text{top-}K}\Big)\to C_{l_{\text{tgt}}}
\end{equation}
where the components are defined as:
\begin{itemize}
    \item $q$: User's query in NL
    \item $D_{l_{\text{src}}}$: Code documentation corpus in source programming language $l_{\text{src}}$
    \item $R(D_{l_{\text{src}}}, q) \to \mathcal{K}$: Retrieve top-$K$ code documents $\mathcal{K}$ related to $q$ from $D_{l_{\text{src}}}$
    \item $p(q, l_{\text{tgt}})$: Task prompt combining query $q$ and target programming language $l_{\text{tgt}}$ specification
    \item $G(p, \mathcal{K}) \to C_{l_{\text{tgt}}}$: Generate code in $l_{\text{tgt}}$ using prompt and retrieved knowledge from $\mathcal{K}$
\end{itemize}

\subsection{Study Settings}

\begin{figure}[t]
\centering
\includegraphics[width=1.0\columnwidth]{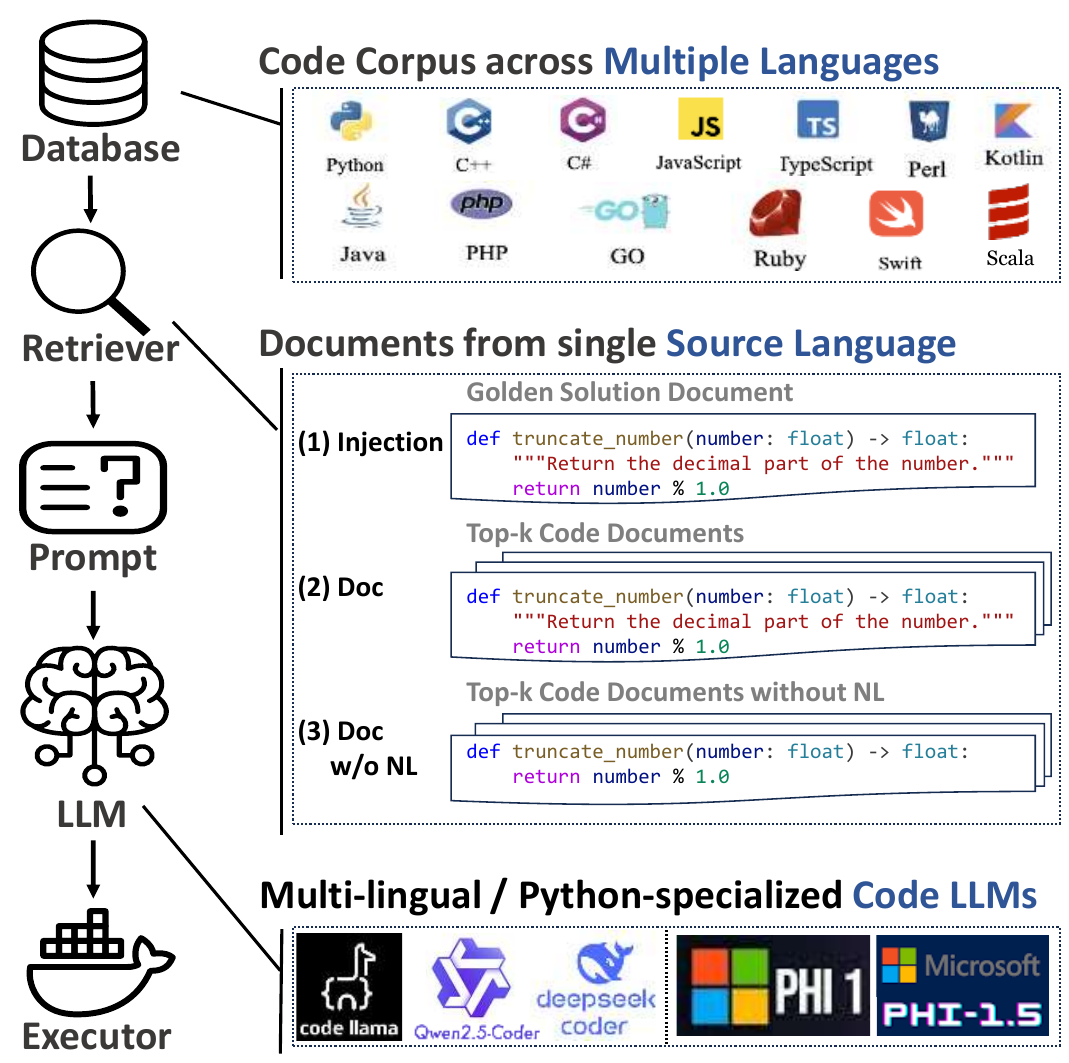}
\caption{RACG pipeline construction and three retrieval experimental settings for cross-lingual knowledge transfer in RACG}
\label{fig:Experimental_setting}
\end{figure}

To investigate cross-lingual knowledge transfer in RACG, we conduct the study across three retrieval experimental settings as shown in Figure~\ref{fig:Experimental_setting}.

(1) \textbf{Golden Solution Document}. We simulate an oracle retriever to isolate the generation, where code LLMs are guaranteed to receive the most relevant code snippets from the specific source \textit{PL} corpora $D_{l_{\text{src}}}$, which can be regarded as a direct knowledge injection means. This injection setup enables us to explore cross-lingual knowledge transfer during generation, independent of retrieval imperfections. In other words, this is also the upper bound of the cross-language knowledge transfer effect in RACG. Specifically, we examine whether code LLMs can leverage knowledge from retrieved code snippets in the source \textit{PL} $l_{\text{src}}$, even when the target generation \textit{PL} $l_{\text{tgt}}$ differs from the $l_{\text{src}}$.
 
(2) \textbf{Top-k Retrieved Code Documents}. 
We implement an end-to-end complete RACG pipeline including multi-lingual code retrievers to evaluate transfer performance with practical retrieval. Top-$K$ relevant code snippets selected by retrievers are provided to LLMs in order to generate better solutions for the code generation task. This setting assesses knowledge transfer performance of code LLMs across \textit{PL}s when the retrieval process is not oracle-based (\eg potential mismatches in programming paradigms or API conventions between retrieved and target \textit{PL}s).

(3) \textbf{Top-k Retrieved Code Documents (without NL)}. Code snippets found online are often isolated fragments lacking corresponding comments or NL descriptions. The only difference between this setting \textit{Doc w/o NL} and the setting \textit{Doc} is that the $D_{l_{\text{src}}}$ provided is pure code without NL. This real-world scenario presents challenges for both the retriever and the LLM: the retriever must effectively identify relevant pure code snippets based on NL queries $q$ ($R(D_{l_{\text{src}}}, q) \to \mathcal{K}$), while the LLM needs to leverage these code-only fragments to improve the output quality ($G(p, \mathcal{K}) \to C_{l_{\text{tgt}}}$). The knowledge transfer in such harsh condition deserves investigation.

Furthermore, we analyze two distinct categories of code LLMs: multi-lingual LLMs (trained on multiple \textit{PL}s) to assess their inherent capacity for cross-lingual knowledge fusion and python-specialized LLMs (trained just on Python) to evaluate their adaptability when augmented with cross-lingual retrieved contexts. Through this dual-lens settings, we aim to uncover the relationship between LLM types (multi-lingual and python-specialized) and cross-lingual transfer efficiency.

\subsection{Experimental Details}

\paragraph{LLMs for Evaluation.}
RACG has two phases: retrieval and generation.
In the retrieval phase, we have oracle retrieval and practical model retrieval.
In the generation phase, we evaluate 5 LLMs, including three representative approximately 7B parameter instruction-tuned multi-lingual LLMs (\textit{CodeLlama-7B-Instruct}~\cite{roziere2023codellama}, \textit{Deepseek-Coder-6.7B-Instruct}~\cite{guo2024deepseek}, \textit{Qwen2.5-Coder-7B-Instruct}~\cite{hui2024qwen2_5coder}) and two python-specialized LLMs (\textit{Phi-1}~\cite{gunasekar2023textbooks} and \textit{Phi-1.5}~\cite{li2023textbooksphi}).
For multi-lingual/python-specialized LLMs, we evaluate and report their averaged performance. Appendix~\ref{sec:model_size} presents the experimental results of a larger code LLM.

\paragraph{Parameter Configuration.}
To ensure reproducibility, we implement greedy decoding with the temperature {0.0} during inference to generate the most probable responses~\cite{chen2025llmsmeetapidocumentation}. The evaluation metric adopts Pass@1 (\ie K=1) under the deterministic setting. Additionally, we report the results of Pass@5 on the data subset in Appendix~\ref{sec:pass_at_5}.
We evaluate the LLMs by using a unified prompt template to instruct LLMs to utilize the code corpus, thereby enhancing code generation performance, as illustrated in Appendix~\ref{sec:prompt}, following the design in papers~\cite{wang2024coderagbench, athiwaratkun2022multilingual}.


\paragraph{Evaluation Metrics.}
Following the metrics of HumanEval-X and most code generation work, we use Pass@K~\cite{chen2021evaluating} to evaluate the correctness of code generated. We quantify the benefits of different corpus by the increase in Pass@K values, which also reflects the knowledge transfer effect across \textit{PL}s in RACG.

\begin{table}[t]
\centering
\resizebox{\columnwidth}{!}{
\setlength{\tabcolsep}{2pt}
    \begin{tabular}{cc|ccccc|c}
    \toprule
    \multicolumn{2}{c|}{\textbf{Multi-lingual LLMs}} & \multicolumn{5}{c|}{\textbf{Target Programming Language }} & \multirow{2}[2]{*}{\textbf{Mean}} \\
    \multicolumn{2}{c|}{\textbf{Pass@k}} & C++   & Go    & Java  & JS    & Python &       \\
    \midrule
    \multicolumn{2}{c|}{\textbf{Baseline (without injection)}} & 54.27  & 42.68  & 61.79  & 58.33  & 59.35  & 55.28 \\
    \midrule
    \multicolumn{1}{c}{\multirow{5}[2]{*}{\textbf{\makecell{Source\\Programming\\Language of\\Corpus}}}} & C++   & \textbackslash{} & \cellcolor[rgb]{ .961,  .98,  1} +4.47 & \cellcolor[rgb]{ .663,  .831,  1} +20.33 & \cellcolor[rgb]{ .69,  .847,  1} +18.90 & \cellcolor[rgb]{ .761,  .882,  1} +15.04 & +14.68 \\
          & Go    & \cellcolor[rgb]{ .871,  .937,  1} +9.15 & \textbackslash{} & \cellcolor[rgb]{ .769,  .886,  1} +14.63 & \cellcolor[rgb]{ .647,  .824,  1} +21.14 & \cellcolor[rgb]{ .737,  .871,  1} +16.26 & +15.29 \\
          & Java  & \cellcolor[rgb]{ .882,  .941,  1} +8.54 & \cellcolor[rgb]{ .812,  .906,  1} +12.40 & \textbackslash{} & \cellcolor[rgb]{ .6,  .8,  1} +23.58 & \cellcolor[rgb]{ .698,  .851,  1} +18.50 & +15.75 \\
          & JS    & \cellcolor[rgb]{ .788,  .894,  1} +13.62 & \cellcolor[rgb]{ .953,  .976,  1} +4.88 & \cellcolor[rgb]{ .831,  .918,  1} +11.38 & \textbackslash{} & \cellcolor[rgb]{ .8,  .902,  1} +13.01 & +10.72 \\
          & Python & +2.24 & \cellcolor[rgb]{ .941,  .973,  1} +5.49 & \cellcolor[rgb]{ .871,  .937,  1} +9.15 & \cellcolor[rgb]{ .78,  .89,  1} +14.02 & \textbackslash{} & +7.72 \\
    \midrule
    \multicolumn{2}{c|}{\textbf{Mean}} & +8.38 & +6.81 & +13.87 & +19.41 & +15.70 & +12.84 \\
    \midrule
    \addlinespace
    \midrule
    \multicolumn{2}{c|}{\textbf{Python-specialized LLMs}} & \multicolumn{5}{c|}{\textbf{Target Programming Language }} & \multirow{2}[2]{*}{\textbf{Mean}} \\
    \multicolumn{2}{c|}{\textbf{Pass@k}} & C++   & Go    & Java  & JS    & Python &       \\
    \midrule
    \multicolumn{2}{c|}{\textbf{Baseline (without injection)}} & 15.55  & 8.54  & 17.38  & 19.51  & 39.63  & 20.12  \\
    \midrule
    \multicolumn{1}{c}{\multirow{5}[2]{*}{\textbf{\makecell{Source\\Programming\\Language of\\Corpus}}}} & C++   & \textbackslash{} & \cellcolor[rgb]{ .922,  .961,  1} +2.74 & \cellcolor[rgb]{ 1,  .91,  .91} -2.13 & \cellcolor[rgb]{ .682,  .843,  1} +10.67 & \cellcolor[rgb]{ 1,  .886,  .886} -2.74 & +2.13 \\
          & Go    & \cellcolor[rgb]{ .839,  .922,  1} +5.49 & \textbackslash{} & \cellcolor[rgb]{ 1,  .949,  .949} -1.22 & \cellcolor[rgb]{ .894,  .949,  1} +3.66 & \cellcolor[rgb]{ 1,  .702,  .702} -7.32 & +0.15 \\
          & Java  & \cellcolor[rgb]{ .757,  .878,  1} +8.23 & \cellcolor[rgb]{ .776,  .89,  1} +7.62 & \textbackslash{} & \cellcolor[rgb]{ .6,  .8,  1} +13.41 & \cellcolor[rgb]{ .929,  .965,  1} +2.44 & +7.93 \\
          & JS    & \cellcolor[rgb]{ .855,  .929,  1} +4.88 & \cellcolor[rgb]{ .875,  .937,  1} +4.27 & 0     & \textbackslash{} & \cellcolor[rgb]{ 1,  .851,  .851} -3.66 & +1.37 \\
          & Python & \cellcolor[rgb]{ .902,  .953,  1} +3.35 & \cellcolor[rgb]{ .937,  .969,  1} +2.13 & \cellcolor[rgb]{ .965,  .984,  1} +1.22 & \cellcolor[rgb]{ .757,  .878,  1} +8.23 & \textbackslash{} & +3.73 \\
    \midrule
    \multicolumn{2}{c|}{\textbf{Mean}} & +5.49 & +4.19 & -0.53 & +8.99 & -2.82 & +3.06 \\
    \bottomrule
    \end{tabular}%
    }%
    \setlength{\belowcaptionskip}{-10pt}
    \caption{Baseline (without injection) and Cross-lingual direct injection performance of multi-lingual LLMs and python-specialized LLMs across different source and target programming languages on \textit{Small but Parallel Multilingual Code Dataset}.}
    \label{tab:pass@k_cross_enhance_ideal_small}
\end{table}

\section{RQ1. How effective is RACG with direct oracle cross-\textit{PL} code injection?}

\subsection{Dataset and Setup}
To study generation mechanisms under \textit{controlled direct knowledge injection}, we use HumanEval-X~\cite{zheng2023codegeex} as \textit{Small but Parallel Multilingual Code Dataset}, a parallel multi-lingual code dataset containing 164 programming problems with verified reference solutions aligned across 5 \textit{PL}s (Python, Java, JavaScript, C++ and Go). This parallel structure eliminates cross-language corpus inconsistencies, enabling rigorous variable control.

For knowledge injection on the \textit{Small but Parallel Multilingual Code Dataset}, we directly include one canonical solution from the source \textit{PL} into the prompt to help code LLMs try to generate solutions in the target \textit{PL}. In this oracle retrieval setup, code LLMs directly access golden source \textit{PL} solutions from the dataset to isolate the influence of the retriever. The aligned problems and solutions ensure fair evaluation across \textit{PL}s.

\subsection{Cross-\textit{PL} Knowledge Injection in RACG Is Non-Trivial}

Table~\ref{tab:pass@k_cross_enhance_ideal_small} shows the experiment results of baseline (without injection) and cross-\textit{PL} direct injection performance of multi-lingual LLMs and python-specialized LLMs across different source and target \textit{PL}s on \textit{Small but Parallel Multilingual Code Dataset}. Knowledge transfer in RACG across \textit{PL}s is not simple and worth exploration.

\begin{mdframed}[style=MyFrame]
\textbf{Finding 1:} Direct injection can enable knowledge transfer across \textit{PL}s.
\end{mdframed}

From cross-lingual knowledge direct injection results in the Table~\ref{tab:pass@k_cross_enhance_ideal_small}, compared to the Baseline (without injection), both the average Pass@K improvement of multi-lingual LLMs ({+12.84\%}) and python-specialized LLMs ({+3.06\%}) show positive values, indicating that the cross-lingual knowledge generally enhances code generation capabilities across multiple \textit{PL}s. Through direct injection, code LLMs can transfer the knowledge across \textit{PL}s in general. This is reasonable because providing reference solutions from other \textit{PL}s has a positive effect on model generation overall.

\begin{mdframed}[style=MyFrame]
\textbf{Finding 2:} For Python-specialized LLMs, direct injection of cross-language knowledge is not always effective.
\end{mdframed}

According to Table~\ref{tab:pass@k_cross_enhance_ideal_small}, the enhancement effects exhibit significant variations between the two distinct types of LLMs. Python-specialized LLMs ({+3.06\%}) demonstrate much smaller improvements compared to multi-lingual LLMs ({+12.84\%}), and even show performance degradation in several \textit{PL} pairs (\eg C++→Java). This is because python-specialized LLMs have a weaker ability to utilize knowledge from other \textit{PL}s, which can instead introduce interference.

\begin{mdframed}[style=MyFrame]
\textbf{Finding 3:} Oracle retrieval does not mean transfer cross-language knowledge without loss.
\end{mdframed}

In this experimental setting of \textit{Direct Knowledge Injection}, we assume that the retrieval process is oracle and the corpus knowledge is complete, representing the upper bound of effectiveness for cross-lingual knowledge transfer. However, even under these ideal conditions, the final generative performance of multi-lingual LLMs (55.28\% + 12.84\% in Table~\ref{tab:pass@k_cross_enhance_ideal_small}) does not exceed 70\%, indicating a significant gap from lossless knowledge transfer (100\% performance). This demonstrates that knowledge transfer across \textit{PL}s through RACG is not straightforward and warrants further exploration.
\begin{table*}[htb]
\centering
\resizebox{\textwidth}{!}{
\setlength{\tabcolsep}{2pt}
    \begin{tabular}{cc|ccccccccccccc|c}
    \toprule
    \multicolumn{2}{c|}{\textbf{Multi-lingual LLMs}} & \multicolumn{13}{c|}{\textbf{Target Programming Language}}                                            & \multirow{2}[2]{*}{\textbf{Mean}} \\
    \multicolumn{2}{c|}{\textbf{Pass@k}} & C++   & C\#   & Go    & Java  & JS    & Kotlin & Perl  & PHP   & Python & Ruby  & Scala & Swift & TS    &       \\
    \midrule
    \multicolumn{2}{c|}{\textbf{Baseline (without RACG)}} & 60.21  & 61.62  & 51.42  & 60.49  & 62.19  & 55.15  & 51.39  & 59.15  & 59.05  & 58.69  & 57.97  & 52.61  & 59.79  & 57.67 \\
    \midrule
    \multicolumn{1}{c}{\multirow{13}[2]{*}{\textbf{\makecell{Source\\Programming\\Language of\\Corpus}}}} & C++   & \textbackslash{} & \cellcolor[rgb]{ .776,  .89,  1} +15.87 & \cellcolor[rgb]{ .757,  .878,  1} +17.46 & \cellcolor[rgb]{ .706,  .855,  1} +20.87 & \cellcolor[rgb]{ .694,  .847,  1} +21.78 & \cellcolor[rgb]{ .784,  .894,  1} +15.41 & \cellcolor[rgb]{ .796,  .898,  1} +14.60 & \cellcolor[rgb]{ .859,  .929,  1} +10.25 & \cellcolor[rgb]{ .863,  .933,  1} +9.85 & \cellcolor[rgb]{ .804,  .902,  1} +13.93 & \cellcolor[rgb]{ .863,  .933,  1} +9.96 & \cellcolor[rgb]{ .827,  .914,  1} +12.38 & \cellcolor[rgb]{ .737,  .871,  1} +18.59 & +15.08 \\
          & C\#   & \cellcolor[rgb]{ .737,  .871,  1} +18.69 & \textbackslash{} & \cellcolor[rgb]{ .784,  .894,  1} +15.28 & \cellcolor[rgb]{ .737,  .871,  1} +18.59 & \cellcolor[rgb]{ .753,  .878,  1} +17.69 & \cellcolor[rgb]{ .706,  .855,  1} +21.01 & \cellcolor[rgb]{ .769,  .886,  1} +16.42 & \cellcolor[rgb]{ .871,  .937,  1} +9.42 & \cellcolor[rgb]{ .89,  .945,  1} +7.90 & \cellcolor[rgb]{ .78,  .89,  1} +15.80 & \cellcolor[rgb]{ .804,  .902,  1} +14.10 & \cellcolor[rgb]{ .792,  .898,  1} +14.80 & \cellcolor[rgb]{ .749,  .875,  1} +17.76 & +15.62 \\
          & Go    & \cellcolor[rgb]{ .729,  .867,  1} +19.30 & \cellcolor[rgb]{ .745,  .875,  1} +18.06 & \textbackslash{} & \cellcolor[rgb]{ .733,  .867,  1} +18.99 & \cellcolor[rgb]{ .788,  .894,  1} +15.13 & \cellcolor[rgb]{ .773,  .886,  1} +16.15 & \cellcolor[rgb]{ .784,  .894,  1} +15.40 & \cellcolor[rgb]{ .855,  .929,  1} +10.48 & \cellcolor[rgb]{ .886,  .945,  1} +8.18 & \cellcolor[rgb]{ .8,  .902,  1} +14.32 & \cellcolor[rgb]{ .816,  .91,  1} +13.16 & \cellcolor[rgb]{ .867,  .933,  1} +9.68 & \cellcolor[rgb]{ .792,  .898,  1} +14.84 & +14.47 \\
          & Java  & \cellcolor[rgb]{ .737,  .871,  1} +18.79 & \cellcolor[rgb]{ .733,  .867,  1} +19.01 & \cellcolor[rgb]{ .796,  .898,  1} +14.66 & \textbackslash{} & \cellcolor[rgb]{ .71,  .855,  1} +20.54 & \cellcolor[rgb]{ .737,  .871,  1} +18.77 & \cellcolor[rgb]{ .71,  .855,  1} +20.54 & \cellcolor[rgb]{ .824,  .914,  1} +12.73 & \cellcolor[rgb]{ .878,  .941,  1} +8.86 & \cellcolor[rgb]{ .796,  .898,  1} +14.66 & \cellcolor[rgb]{ .827,  .914,  1} +12.43 & \cellcolor[rgb]{ .804,  .902,  1} +14.06 & \cellcolor[rgb]{ .749,  .875,  1} +17.90 & +16.08 \\
          & JS    & \cellcolor[rgb]{ .847,  .925,  1} +11.08 & \cellcolor[rgb]{ .8,  .902,  1} +14.16 & \cellcolor[rgb]{ .875,  .937,  1} +9.10 & \cellcolor[rgb]{ .882,  .941,  1} +8.40 & \textbackslash{} & \cellcolor[rgb]{ .824,  .914,  1} +12.64 & \cellcolor[rgb]{ .839,  .922,  1} +11.55 & \cellcolor[rgb]{ .91,  .957,  1} +6.42 & \cellcolor[rgb]{ .894,  .949,  1} +7.62 & \cellcolor[rgb]{ .804,  .902,  1} +14.14 & \cellcolor[rgb]{ .922,  .961,  1} +5.63 & \cellcolor[rgb]{ .914,  .957,  1} +6.12 & \cellcolor[rgb]{ .6,  .8,  1} +28.30 & +11.26 \\
          & Kotlin & \cellcolor[rgb]{ .808,  .906,  1} +13.87 & \cellcolor[rgb]{ .749,  .875,  1} +17.84 & \cellcolor[rgb]{ .863,  .933,  1} +9.76 & \cellcolor[rgb]{ .8,  .902,  1} +14.37 & \cellcolor[rgb]{ .761,  .882,  1} +17.19 & \textbackslash{} & \cellcolor[rgb]{ .824,  .914,  1} +12.75 & \cellcolor[rgb]{ .878,  .941,  1} +8.67 & \cellcolor[rgb]{ .871,  .937,  1} +9.37 & \cellcolor[rgb]{ .733,  .867,  1} +18.89 & \cellcolor[rgb]{ .91,  .957,  1} +6.64 & \cellcolor[rgb]{ .933,  .969,  1} +4.88 & \cellcolor[rgb]{ .773,  .886,  1} +16.36 & +12.55 \\
          & Perl  & \cellcolor[rgb]{ .847,  .925,  1} +11.05 & \cellcolor[rgb]{ .769,  .886,  1} +16.51 & \cellcolor[rgb]{ .824,  .914,  1} +12.74 & \cellcolor[rgb]{ .745,  .875,  1} +18.06 & \cellcolor[rgb]{ .871,  .937,  1} +9.18 & \cellcolor[rgb]{ .788,  .894,  1} +15.19 & \textbackslash{} & \cellcolor[rgb]{ .863,  .933,  1} +9.79 & \cellcolor[rgb]{ .812,  .906,  1} +13.40 & \cellcolor[rgb]{ .706,  .855,  1} +20.91 & \cellcolor[rgb]{ .863,  .933,  1} +9.84 & \cellcolor[rgb]{ .839,  .922,  1} +11.64 & \cellcolor[rgb]{ .824,  .914,  1} +12.49 & +13.40 \\
          & PHP   & \cellcolor[rgb]{ .941,  .973,  1} +4.21 & \cellcolor[rgb]{ .933,  .969,  1} +4.82 & \cellcolor[rgb]{ .969,  .984,  1} +2.32 & \cellcolor[rgb]{ .922,  .961,  1} +5.59 & \cellcolor[rgb]{ .957,  .98,  1} +3.18 & \cellcolor[rgb]{ .98,  .992,  1} +1.56 & \cellcolor[rgb]{ .949,  .976,  1} +3.69 & \textbackslash{} & \cellcolor[rgb]{ .961,  .98,  1} +2.99 & \cellcolor[rgb]{ .906,  .953,  1} +6.92 & \cellcolor[rgb]{ 1,  .788,  .788} -1.20 & +0.08 & \cellcolor[rgb]{ .918,  .961,  1} +5.96 & +3.34 \\
          & Python & \cellcolor[rgb]{ .894,  .949,  1} +7.58 & \cellcolor[rgb]{ .8,  .902,  1} +14.28 & \cellcolor[rgb]{ .867,  .933,  1} +9.50 & \cellcolor[rgb]{ .808,  .906,  1} +13.84 & \cellcolor[rgb]{ .957,  .98,  1} +3.06 & \cellcolor[rgb]{ .859,  .929,  1} +10.08 & \cellcolor[rgb]{ .851,  .925,  1} +10.72 & \cellcolor[rgb]{ .91,  .957,  1} +6.42 & \textbackslash{} & \cellcolor[rgb]{ .902,  .953,  1} +6.98 & \cellcolor[rgb]{ .898,  .949,  1} +7.24 & \cellcolor[rgb]{ .91,  .957,  1} +6.48 & \cellcolor[rgb]{ .953,  .976,  1} +3.35 & +8.29 \\
          & Ruby  & \cellcolor[rgb]{ 1,  .969,  .969} -0.16 & \cellcolor[rgb]{ .8,  .902,  1} +14.38 & \cellcolor[rgb]{ .906,  .953,  1} +6.81 & \cellcolor[rgb]{ .875,  .937,  1} +9.10 & \cellcolor[rgb]{ .984,  .992,  1} +1.21 & \cellcolor[rgb]{ .808,  .906,  1} +13.60 & \cellcolor[rgb]{ .898,  .949,  1} +7.27 & \cellcolor[rgb]{ .933,  .969,  1} +4.87 & \cellcolor[rgb]{ .973,  .988,  1} +2.03 & \textbackslash{} & \cellcolor[rgb]{ .898,  .949,  1} +7.37 & \cellcolor[rgb]{ .906,  .953,  1} +6.73 & \cellcolor[rgb]{ .925,  .965,  1} +5.30 & +6.54 \\
          & Scala & \cellcolor[rgb]{ .8,  .902,  1} +14.42 & \cellcolor[rgb]{ .78,  .89,  1} +15.78 & \cellcolor[rgb]{ .871,  .937,  1} +9.35 & \cellcolor[rgb]{ .863,  .933,  1} +9.98 & \cellcolor[rgb]{ .867,  .933,  1} +9.71 & \cellcolor[rgb]{ 1,  .702,  .702} -1.71 & \cellcolor[rgb]{ .843,  .922,  1} +11.36 & \cellcolor[rgb]{ .906,  .953,  1} +6.88 & \cellcolor[rgb]{ .898,  .949,  1} +7.45 & \cellcolor[rgb]{ .765,  .882,  1} +16.74 & \textbackslash{} & \cellcolor[rgb]{ .933,  .969,  1} +4.95 & \cellcolor[rgb]{ .855,  .929,  1} +10.46 & +9.61 \\
          & Swift & \cellcolor[rgb]{ .776,  .89,  1} +16.09 & \cellcolor[rgb]{ .765,  .882,  1} +16.86 & \cellcolor[rgb]{ .855,  .929,  1} +10.35 & \cellcolor[rgb]{ .776,  .89,  1} +15.92 & \cellcolor[rgb]{ .776,  .89,  1} +15.95 & \cellcolor[rgb]{ .824,  .914,  1} +12.70 & \cellcolor[rgb]{ .808,  .906,  1} +13.80 & \cellcolor[rgb]{ .871,  .937,  1} +9.33 & \cellcolor[rgb]{ .851,  .925,  1} +10.66 & \cellcolor[rgb]{ .78,  .89,  1} +15.80 & \cellcolor[rgb]{ .843,  .922,  1} +11.35 & \textbackslash{} & \cellcolor[rgb]{ .784,  .894,  1} +15.32 & +13.68 \\
          & TS    & \cellcolor[rgb]{ .792,  .898,  1} +14.87 & \cellcolor[rgb]{ .827,  .914,  1} +12.48 & \cellcolor[rgb]{ .859,  .929,  1} +10.02 & \cellcolor[rgb]{ .824,  .914,  1} +12.53 & \cellcolor[rgb]{ .718,  .859,  1} +20.04 & \cellcolor[rgb]{ .835,  .918,  1} +11.67 & \cellcolor[rgb]{ .871,  .937,  1} +9.30 & \cellcolor[rgb]{ .804,  .902,  1} +13.99 & \cellcolor[rgb]{ .882,  .941,  1} +8.38 & \cellcolor[rgb]{ .82,  .91,  1} +12.99 & \cellcolor[rgb]{ .918,  .961,  1} +5.98 & \cellcolor[rgb]{ .882,  .941,  1} +8.51 & \textbackslash{} & +11.73 \\
    \midrule
    \multicolumn{2}{c|}{\textbf{Mean}} & +12.48 & +15.00 & +10.61 & +13.85 & +12.89 & +12.26 & +12.28 & +9.10 & +8.06 & +14.34 & +8.54 & +8.36 & +13.89 & +11.67 \\
    \midrule
    \addlinespace
    \midrule
    \multicolumn{2}{c|}{\textbf{Python-specialized LLMs}} & \multicolumn{13}{c|}{\textbf{Target Programming Language}}                                            & \multirow{2}[2]{*}{\textbf{Mean}} \\
    \multicolumn{2}{c|}{\textbf{Pass@k}} & C++   & C\#   & Go    & Java  & JS    & Kotlin & Perl  & PHP   & Python & Ruby  & Scala & Swift & TS    &       \\
    \midrule
    \multicolumn{2}{c|}{\textbf{Baseline (without RACG)}} & 23.65  & 25.29  & 7.73  & 21.07  & 26.48  & 14.85  & 10.54  & 24.31  & 37.85  & 23.25  & 15.75  & 14.78  & 25.56  & 20.85 \\
    \midrule
    \multicolumn{1}{c}{\multirow{13}[2]{*}{\textbf{\makecell{Source\\Programming\\Language of\\Corpus}}}} & C++   & \textbackslash{} & \cellcolor[rgb]{ .886,  .945,  1} +13.04 & \cellcolor[rgb]{ .961,  .98,  1} +4.54 & \cellcolor[rgb]{ .925,  .965,  1} +8.52 & \cellcolor[rgb]{ .898,  .949,  1} +11.60 & \cellcolor[rgb]{ .969,  .984,  1} +3.64 & \cellcolor[rgb]{ .969,  .984,  1} +3.60 & \cellcolor[rgb]{ .859,  .929,  1} +16.06 & \cellcolor[rgb]{ 1,  .973,  .973} -1.14 & \cellcolor[rgb]{ .945,  .973,  1} +6.26 & \cellcolor[rgb]{ .992,  .996,  1} +1.09 & \cellcolor[rgb]{ .98,  .992,  1} +2.35 & \cellcolor[rgb]{ .902,  .953,  1} +11.30 & +6.74 \\
          & C\#   & \cellcolor[rgb]{ .949,  .976,  1} +6.02 & \textbackslash{} & \cellcolor[rgb]{ .973,  .988,  1} +3.32 & \cellcolor[rgb]{ .992,  .996,  1} +1.05 & \cellcolor[rgb]{ .988,  .996,  1} +1.59 & \cellcolor[rgb]{ .969,  .984,  1} +3.87 & \cellcolor[rgb]{ .984,  .992,  1} +2.03 & \cellcolor[rgb]{ .937,  .969,  1} +7.30 & \cellcolor[rgb]{ 1,  .902,  .902} -4.36 & \cellcolor[rgb]{ .961,  .98,  1} +4.63 & \cellcolor[rgb]{ 1,  .965,  .965} -1.47 & \cellcolor[rgb]{ .988,  .996,  1} +1.50 & \cellcolor[rgb]{ 1,  .973,  .973} -1.16 & +2.03 \\
          & Go    & \cellcolor[rgb]{ 1,  .988,  .988} -0.48 & \cellcolor[rgb]{ .969,  .984,  1} +4.00 & \textbackslash{} & \cellcolor[rgb]{ 1,  .98,  .98} -0.75 & \cellcolor[rgb]{ .996,  1,  1} +0.84 & \cellcolor[rgb]{ 1,  .902,  .902} -4.35 & \cellcolor[rgb]{ .984,  .992,  1} +1.89 & \cellcolor[rgb]{ .945,  .973,  1} +6.61 & \cellcolor[rgb]{ 1,  .98,  .98} -0.72 & \cellcolor[rgb]{ 1,  .875,  .875} -5.57 & \cellcolor[rgb]{ 1,  .875,  .875} -5.55 & \cellcolor[rgb]{ 1,  .859,  .859} -6.40 & \cellcolor[rgb]{ 1,  .906,  .906} -4.21 & -1.22 \\
          & Java  & \cellcolor[rgb]{ .953,  .976,  1} +5.40 & \cellcolor[rgb]{ .969,  .984,  1} +4.00 & \cellcolor[rgb]{ .98,  .992,  1} +2.66 & \textbackslash{} & \cellcolor[rgb]{ 1,  .957,  .957} -1.94 & \cellcolor[rgb]{ .992,  .996,  1} +0.98 & \cellcolor[rgb]{ .976,  .988,  1} +2.72 & \cellcolor[rgb]{ .886,  .945,  1} +13.17 & \cellcolor[rgb]{ 1,  .788,  .788} -9.53 & \cellcolor[rgb]{ .992,  .996,  1} +1.05 & \cellcolor[rgb]{ .996,  1,  1} +0.57 & +0.32 & \cellcolor[rgb]{ .988,  .996,  1} +1.72 & +1.76 \\
          & JS    & \cellcolor[rgb]{ .91,  .957,  1} +10.26 & \cellcolor[rgb]{ .929,  .965,  1} +8.38 & \cellcolor[rgb]{ .976,  .988,  1} +2.88 & \cellcolor[rgb]{ .925,  .965,  1} +8.74 & \textbackslash{} & \cellcolor[rgb]{ .973,  .988,  1} +3.12 & \cellcolor[rgb]{ .969,  .984,  1} +3.65 & \cellcolor[rgb]{ .898,  .949,  1} +11.61 & \cellcolor[rgb]{ 1,  .898,  .898} -4.57 & \cellcolor[rgb]{ .953,  .976,  1} +5.44 & \cellcolor[rgb]{ .973,  .988,  1} +3.18 & \cellcolor[rgb]{ .988,  .996,  1} +1.39 & \cellcolor[rgb]{ .6,  .8,  1} +45.40 & +8.29 \\
          & Kotlin & \cellcolor[rgb]{ .996,  1,  1} +0.58 & \cellcolor[rgb]{ .941,  .973,  1} +6.95 & \cellcolor[rgb]{ .973,  .988,  1} +3.21 & \cellcolor[rgb]{ .941,  .973,  1} +6.76 & \cellcolor[rgb]{ .929,  .965,  1} +8.29 & \textbackslash{} & \cellcolor[rgb]{ .992,  .996,  1} +1.11 & \cellcolor[rgb]{ .914,  .957,  1} +10.06 & \cellcolor[rgb]{ 1,  .878,  .878} -5.46 & \cellcolor[rgb]{ 1,  .992,  .992} -0.18 & \cellcolor[rgb]{ .949,  .976,  1} +5.93 & \cellcolor[rgb]{ .969,  .984,  1} +3.63 & \cellcolor[rgb]{ .957,  .98,  1} +5.24 & +3.84 \\
          & Perl  & \cellcolor[rgb]{ 1,  .816,  .816} -8.33 & \cellcolor[rgb]{ .984,  .992,  1} +2.14 & \cellcolor[rgb]{ 1,  .98,  .98} -0.88 & \cellcolor[rgb]{ .976,  .988,  1} +2.94 & \cellcolor[rgb]{ 1,  .816,  .816} -8.39 & \cellcolor[rgb]{ 1,  .859,  .859} -6.35 & \textbackslash{} & \cellcolor[rgb]{ 1,  .941,  .941} -2.63 & \cellcolor[rgb]{ 1,  .953,  .953} -2.12 & +0.10 & \cellcolor[rgb]{ 1,  .855,  .855} -6.45 & \cellcolor[rgb]{ 1,  .843,  .843} -7.15 & \cellcolor[rgb]{ 1,  .816,  .816} -8.25 & -3.78 \\
          & PHP   & \cellcolor[rgb]{ .953,  .976,  1} +5.73 & \cellcolor[rgb]{ .961,  .98,  1} +4.47 & \cellcolor[rgb]{ .996,  1,  1} +0.45 & \cellcolor[rgb]{ .98,  .992,  1} +2.50 & \cellcolor[rgb]{ .984,  .992,  1} +1.94 & \cellcolor[rgb]{ .969,  .984,  1} +3.92 & \cellcolor[rgb]{ .973,  .988,  1} +3.42 & \textbackslash{} & \cellcolor[rgb]{ 1,  .835,  .835} -7.37 & \cellcolor[rgb]{ 1,  .992,  .992} -0.27 & \cellcolor[rgb]{ 1,  .992,  .992} -0.29 & +0.37 & \cellcolor[rgb]{ .984,  .992,  1} +2.02 & +1.41 \\
          & Python & \cellcolor[rgb]{ 1,  .937,  .937} -2.84 & \cellcolor[rgb]{ 1,  .992,  .992} -0.34 & \cellcolor[rgb]{ 1,  .988,  .988} -0.49 & \cellcolor[rgb]{ 1,  .965,  .965} -1.49 & \cellcolor[rgb]{ 1,  .8,  .8} -9.00 & \cellcolor[rgb]{ 1,  .847,  .847} -6.91 & \cellcolor[rgb]{ .969,  .984,  1} +3.92 & \cellcolor[rgb]{ 1,  .933,  .933} -2.89 & \textbackslash{} & \cellcolor[rgb]{ 1,  .925,  .925} -3.35 & \cellcolor[rgb]{ 1,  .902,  .902} -4.46 & \cellcolor[rgb]{ 1,  .816,  .816} -8.27 & \cellcolor[rgb]{ 1,  .78,  .78} -9.88 & -3.83 \\
          & Ruby  & \cellcolor[rgb]{ 1,  .831,  .831} -7.66 & \cellcolor[rgb]{ .91,  .957,  1} +10.61 & \cellcolor[rgb]{ 1,  .973,  .973} -1.16 & \cellcolor[rgb]{ 1,  .941,  .941} -2.54 & \cellcolor[rgb]{ 1,  .882,  .882} -5.34 & \cellcolor[rgb]{ 1,  .855,  .855} -6.59 & \cellcolor[rgb]{ .973,  .988,  1} +3.23 & \cellcolor[rgb]{ 1,  .878,  .878} -5.53 & \cellcolor[rgb]{ 1,  .945,  .945} -2.37 & \textbackslash{} & \cellcolor[rgb]{ 1,  .925,  .925} -3.23 & \cellcolor[rgb]{ 1,  .882,  .882} -5.28 & \cellcolor[rgb]{ 1,  .875,  .875} -5.67 & -2.63 \\
          & Scala & \cellcolor[rgb]{ 1,  .937,  .937} -2.70 & \cellcolor[rgb]{ .976,  .988,  1} +2.71 & \cellcolor[rgb]{ 1,  .965,  .965} -1.60 & \cellcolor[rgb]{ 1,  .98,  .98} -0.75 & \cellcolor[rgb]{ 1,  .91,  .91} -4.06 & \cellcolor[rgb]{ 1,  .702,  .702} -13.59 & \cellcolor[rgb]{ 1,  .992,  .992} -0.24 & \cellcolor[rgb]{ .976,  .988,  1} +2.72 & \cellcolor[rgb]{ 1,  .82,  .82} -8.21 & +0.19 & \textbackslash{} & \cellcolor[rgb]{ 1,  .89,  .89} -4.91 & \cellcolor[rgb]{ 1,  .82,  .82} -8.16 & -3.22 \\
          & Swift & \cellcolor[rgb]{ 1,  .992,  .992} -0.24 & \cellcolor[rgb]{ 1,  .788,  .788} -9.53 & \cellcolor[rgb]{ .988,  .996,  1} +1.61 & \cellcolor[rgb]{ .973,  .988,  1} +3.47 & +0.26 & \cellcolor[rgb]{ 1,  .859,  .859} -6.26 & \cellcolor[rgb]{ .98,  .992,  1} +2.49 & \cellcolor[rgb]{ .925,  .965,  1} +8.72 & \cellcolor[rgb]{ 1,  .871,  .871} -5.89 & +0.14 & \cellcolor[rgb]{ 1,  .847,  .847} -6.88 & \textbackslash{} & \cellcolor[rgb]{ .976,  .988,  1} +2.70 & -0.78 \\
          & TS    & \cellcolor[rgb]{ .922,  .961,  1} +9.30 & \cellcolor[rgb]{ .965,  .984,  1} +4.09 & \cellcolor[rgb]{ .929,  .965,  1} +8.35 & \cellcolor[rgb]{ .957,  .98,  1} +5.22 & \cellcolor[rgb]{ .749,  .875,  1} +28.90 & \cellcolor[rgb]{ .992,  .996,  1} +0.98 & \cellcolor[rgb]{ 1,  .992,  .992} -0.19 & \cellcolor[rgb]{ .969,  .984,  1} +3.84 & \cellcolor[rgb]{ 1,  .929,  .929} -3.05 & \cellcolor[rgb]{ 1,  .996,  .996} -0.09 & \cellcolor[rgb]{ .984,  .992,  1} +1.85 & \cellcolor[rgb]{ 1,  .98,  .98} -0.85 & \textbackslash{} & +4.86 \\
    \midrule
    \multicolumn{2}{c|}{\textbf{Mean}} & +1.25 & +4.21 & +1.91 & +2.81 & +2.06 & -2.30 & +2.30 & +5.75 & -4.57 & +0.70 & -1.31 & -1.94 & +2.59 & +1.04 \\
    \bottomrule
    \end{tabular}%
    }
    \caption{RACG performance compared to baseline (without RACG) when the retrieval corpus and generation task involve different programming languages on \textit{Large Multilingual Code Dataset} in \textit{Doc} setting.}
    \label{tab:pass@k_cross_enhance_ideal_big}
\end{table*}

\section{RQ2. How effective is RACG at cross-\textit{PL} knowledge transfer via code retrieval?}

\subsection{Dataset and Setup}
To support broader language coverage and code documents retrieval in Figure~\ref{fig:Experimental_setting}, we extend existing datasets from previous work~\cite{athiwaratkun2022multilingual, chai2024mceval}, which initially contained partially aligned problems across 13 \textit{PL}s but lacked reference solutions. To ensure reproducibility and solution quality, we employ the powerful open-source LLM, \ie~\textit{Qwen2-72B-Instruct-GPTQ-Int4} to generate missing reference solutions using multi-lingual RACG, followed by verification through unit testing. We conduct five iterations of solution generation with verification in total. Finally, we construct a \textit{Large Multilingual Code Dataset} and unify both the data format and evaluation methodology according to the HumanEval-X~\cite{zheng2023codegeex} benchmark.
Our datasets provide not only test cases but also verified reference solutions, establishing a foundation for constructing a multi-\textit{PL} code generation benchmark and retrieval repository. Following the canonical document setup described in \cite{wang2024coderagbench}, we create code documents for the retrieval corpus (i.e., reference solutions with NL comments).

Through rigorous unit testing and manual review, we ensure the correctness of our data. This data synthesis method does not introduce potential dataset bias; details can be found in the Appendix~\ref{sec:bias}.
This pipeline produces 13910 high-quality datapoints, approximately 1,000 datapoints per language across 13 \textit{PL}s (Python: 1181; Kotlin: 1071; Java: 1139; Ruby: 1103; JavaScript (JS): 1133; PHP: 1158; TypeScript (TS): 1059; C++: 1038; C\#: 1050; Go: 905; Perl: 1082; Scala: 1054; Swift: 937), providing balanced coverage while maintaining high degree of alignment.

For the complete RACG pipeline on \textit{Large Multilingual Code Dataset}, to control for variables, we fix the retrieval window size to {3} due to the context length limitation of Python-specialized LLMs, that is, using the Top-3 relevant code documents for each query. We also report the experiment results of different retrieval window sizes in Appendix~\ref{sec:window_size}.
In the retrieval phase, we employ a state-of-the-art embedding model \textit{CodeRankEmbed}~\cite{suresh2024cornstack} for the code retrieval task, which supports multi-lingual code retrieval. 
In the LLM generation stage, we use a unified prompt template to evaluate as shown in Appendix~\ref{sec:prompt}.

\subsection{Cross-\textit{PL} Knowledge Transfer in RACG Is Unequal.}

From Table~\ref{tab:pass@k_cross_enhance_ideal_big}, we report the RACG performance improvements on \textit{Large Multilingual Code Dataset} in \textit{Doc} setting. RACG enables unequal cross-lingual knowledge transfer, and its efficacy depends on linguistic affinity of \textit{PL} pair and diversity of LLM pretraining corpus.

\begin{mdframed}[style=MyFrame]
\textbf{Finding 4:} Balanced code generation abilities do not mean equal knowledge transfer abilities.
\end{mdframed}

While multi-lingual LLMs show relatively balanced code generation performance across 13 \textit{PL}s (aligning with findings from the work~\cite{cassano2022multipl}), their capacity to leverage retrieved cross-lingual knowledge shows significant disparities. Overall, the results in Table~\ref{tab:pass@k_cross_enhance_ideal_big} show that the enhancement effects vary considerably across different \textit{PL} pairs, with a maximum improvement of +28.30\%, while the use of PHP and Scala corpora even degrades the generation performance for the specific target \textit{PL}.

The mainstream \textit{PL} used for training is not always a suitable knowledge transfer vehicle. While both Python and Java dominate multi-lingual LLMs pretraining, Java demonstrates superior efficacy as a retrieval corpus (Java→others average improvement: {+16.08\%} in comparison with Python→others: {+8.29\%}). This reveals an unexpected difference between training and retrieval.

\begin{mdframed}[style=MyFrame]
\textbf{Finding 5:} Knowledge transfer is always very effective when two \textit{PL}s are very similar.
\end{mdframed}

We also observe linguistic affinity patterns: JavaScript and TypeScript exhibit outstanding bidirectional enhancement effects. When using TypeScript as the corpus, JavaScript shows the most significant improvement ({+20.04\%} and {+28.90\%} in Table~\ref{tab:pass@k_cross_enhance_ideal_big}). Meanwhile, TypeScript demonstrates gains the strongest enhancement ({+28.30\%} and {+45.40\%} in Table~\ref{tab:pass@k_cross_enhance_ideal_big}) when corpus is JavaScript. This is because TypeScript is a superset of JavaScript, and the two are very similar.

\begin{mdframed}[style=MyFrame]
\textbf{Finding 6:} RACG is not good at enabling python-specialized LLMs to transfer knowledge; Diverse training corpus is key to transfer knowledge across \textit{PL}s for LLMs in RACG.
\end{mdframed}

Horizontally, python-specialized LLMs performs poorly when Python is source \textit{PL} because they lack the capability to generate solutions in other \textit{PL}s.
Vertically, python-specialized LLMs improve in several non-specialized \textit{PL}s through cross-lingual RACG but suffer {4.57\%} degradation (Table~\ref{tab:pass@k_cross_enhance_ideal_big}) in their native \textit{PL} tasks (\ie, Python task performance for a Python-specialized LLM), indicating that knowledge in other \textit{PL}s acts as interference because they cannot comprehend and utilize it.. 
Moreover, the overall improvement is minimal, only +1.04\% in Table~\ref{tab:pass@k_cross_enhance_ideal_big}, indicating that knowledge transfer across \textit{PL}s is rather ineffective.

This contrasts with multi-lingual LLMs, which demonstrate all positive gains across 13 \textit{PL}s. Furthermore, under the same \textit{PL}s settings, multi-lingual LLMs universally achieve greater improvements compared to python-specialized LLMs in cross-lingual RACG, suggesting stronger cross-lingual knowledge transfer capabilities derived from their diverse pretraining.
\section{RQ3. How does NL information in code affect cross-\textit{PL} knowledge transfer in RACG?}

\begin{table*}[ht]
\centering
\resizebox{\textwidth}{!}{
\setlength{\tabcolsep}{2pt}
    \begin{tabular}{cc|ccccccccccccc|c}
    \toprule
    \multicolumn{2}{c|}{\textbf{Multi-lingual LLMs}} & \multicolumn{13}{c|}{\textbf{Target Programming Language}}                                            & \multirow{2}[2]{*}{\textbf{Mean}} \\
    \multicolumn{2}{c|}{\textbf{Pass@k}} & C++   & C\#   & Go    & Java  & JS    & Kotlin & Perl  & PHP   & Python & Ruby  & Scala & Swift & TS    &       \\
    \midrule
    \multicolumn{2}{c|}{\textbf{Baseline (without RACG)}} & 60.21  & 61.62  & 51.42  & 60.49  & 62.19  & 55.15  & 51.39  & 59.15  & 59.05  & 58.69  & 57.97  & 52.61  & 59.79  & 57.67 \\
    \midrule
    \multicolumn{1}{c}{\multirow{13}[2]{*}{\textbf{\makecell{Source\\Programming\\Language of\\Corpus}}}} & C++   & \textbackslash{} & \cellcolor[rgb]{ .706,  .855,  1} +14.06 & \cellcolor[rgb]{ .749,  .875,  1} +11.97 & \cellcolor[rgb]{ .702,  .851,  1} +14.11 & \cellcolor[rgb]{ .667,  .835,  1} +15.92 & \cellcolor[rgb]{ .757,  .878,  1} +11.67 & \cellcolor[rgb]{ .804,  .902,  1} +9.40 & \cellcolor[rgb]{ .647,  .824,  1} +16.70 & \cellcolor[rgb]{ .839,  .922,  1} +7.71 & \cellcolor[rgb]{ .761,  .882,  1} +11.45 & \cellcolor[rgb]{ .816,  .91,  1} +8.86 & \cellcolor[rgb]{ .804,  .902,  1} +9.36 & \cellcolor[rgb]{ .741,  .871,  1} +12.34 & +11.96 \\
          & C\#   & \cellcolor[rgb]{ .718,  .859,  1} +13.49 & \textbackslash{} & \cellcolor[rgb]{ .741,  .871,  1} +12.30 & \cellcolor[rgb]{ .694,  .847,  1} +14.57 & \cellcolor[rgb]{ .671,  .835,  1} +15.59 & \cellcolor[rgb]{ .647,  .824,  1} +16.87 & \cellcolor[rgb]{ .812,  .906,  1} +9.03 & \cellcolor[rgb]{ .667,  .835,  1} +15.83 & \cellcolor[rgb]{ .835,  .918,  1} +7.87 & \cellcolor[rgb]{ .757,  .878,  1} +11.51 & \cellcolor[rgb]{ .761,  .882,  1} +11.42 & \cellcolor[rgb]{ .784,  .894,  1} +10.25 & \cellcolor[rgb]{ .725,  .863,  1} +13.06 & +12.65 \\
          & Go    & \cellcolor[rgb]{ .737,  .871,  1} +12.52 & \cellcolor[rgb]{ .725,  .863,  1} +13.08 & \textbackslash{} & \cellcolor[rgb]{ .694,  .847,  1} +14.49 & \cellcolor[rgb]{ .706,  .855,  1} +13.92 & \cellcolor[rgb]{ .729,  .867,  1} +12.82 & \cellcolor[rgb]{ .792,  .898,  1} +9.83 & \cellcolor[rgb]{ .698,  .851,  1} +14.31 & \cellcolor[rgb]{ .808,  .906,  1} +9.12 & \cellcolor[rgb]{ .796,  .898,  1} +9.70 & \cellcolor[rgb]{ .796,  .898,  1} +9.80 & \cellcolor[rgb]{ .851,  .925,  1} +7.15 & \cellcolor[rgb]{ .729,  .867,  1} +12.86 & +11.63 \\
          & Java  & \cellcolor[rgb]{ .698,  .851,  1} +14.32 & \cellcolor[rgb]{ .655,  .827,  1} +16.41 & \cellcolor[rgb]{ .78,  .89,  1} +10.42 & \textbackslash{} & \cellcolor[rgb]{ .659,  .831,  1} +16.21 & \cellcolor[rgb]{ .678,  .839,  1} +15.38 & \cellcolor[rgb]{ .702,  .851,  1} +14.11 & \cellcolor[rgb]{ .6,  .8,  1} +18.91 & \cellcolor[rgb]{ .831,  .918,  1} +8.10 & \cellcolor[rgb]{ .753,  .878,  1} +11.82 & \cellcolor[rgb]{ .776,  .89,  1} +10.59 & \cellcolor[rgb]{ .753,  .878,  1} +11.70 & \cellcolor[rgb]{ .741,  .871,  1} +12.26 & +13.35 \\
          & JS    & \cellcolor[rgb]{ .835,  .918,  1} +7.87 & \cellcolor[rgb]{ .765,  .882,  1} +11.21 & \cellcolor[rgb]{ .886,  .945,  1} +5.56 & \cellcolor[rgb]{ .843,  .922,  1} +7.46 & \textbackslash{} & \cellcolor[rgb]{ .788,  .894,  1} +10.15 & \cellcolor[rgb]{ .847,  .925,  1} +7.30 & \cellcolor[rgb]{ .804,  .902,  1} +9.38 & \cellcolor[rgb]{ .882,  .941,  1} +5.67 & \cellcolor[rgb]{ .804,  .902,  1} +9.40 & \cellcolor[rgb]{ .843,  .922,  1} +7.43 & \cellcolor[rgb]{ .941,  .973,  1} +2.95 & \cellcolor[rgb]{ .616,  .808,  1} +18.18 & +8.55 \\
          & Kotlin & \cellcolor[rgb]{ .773,  .886,  1} +10.89 & \cellcolor[rgb]{ .686,  .843,  1} +14.92 & \cellcolor[rgb]{ .863,  .933,  1} +6.52 & \cellcolor[rgb]{ .765,  .882,  1} +11.30 & \cellcolor[rgb]{ .71,  .855,  1} +13.77 & \textbackslash{} & \cellcolor[rgb]{ .871,  .937,  1} +6.13 & \cellcolor[rgb]{ .745,  .875,  1} +12.12 & \cellcolor[rgb]{ .843,  .922,  1} +7.56 & \cellcolor[rgb]{ .706,  .855,  1} +14.02 & \cellcolor[rgb]{ 1,  .702,  .702} -8.29 & \cellcolor[rgb]{ 1,  .839,  .839} -4.41 & \cellcolor[rgb]{ .773,  .886,  1} +10.82 & +7.95 \\
          & Perl  & \cellcolor[rgb]{ .741,  .871,  1} +12.30 & \cellcolor[rgb]{ .765,  .882,  1} +11.24 & \cellcolor[rgb]{ .784,  .894,  1} +10.24 & \cellcolor[rgb]{ .694,  .847,  1} +14.63 & \cellcolor[rgb]{ .643,  .824,  1} +17.03 & \cellcolor[rgb]{ .749,  .875,  1} +11.89 & \textbackslash{} & \cellcolor[rgb]{ .725,  .863,  1} +12.98 & \cellcolor[rgb]{ .765,  .882,  1} +11.21 & \cellcolor[rgb]{ .675,  .839,  1} +15.53 & \cellcolor[rgb]{ .804,  .902,  1} +9.39 & \cellcolor[rgb]{ .843,  .922,  1} +7.44 & \cellcolor[rgb]{ .686,  .843,  1} +14.89 & +12.40 \\
          & PHP   & \cellcolor[rgb]{ .969,  .984,  1} +1.57 & \cellcolor[rgb]{ .929,  .965,  1} +3.37 & \cellcolor[rgb]{ .965,  .984,  1} +1.80 & \cellcolor[rgb]{ .894,  .949,  1} +5.15 & \cellcolor[rgb]{ .914,  .957,  1} +4.15 & \cellcolor[rgb]{ .941,  .973,  1} +2.89 & +0.18 & \textbackslash{} & \cellcolor[rgb]{ .957,  .98,  1} +2.20 & \cellcolor[rgb]{ .925,  .965,  1} +3.60 & \cellcolor[rgb]{ .941,  .973,  1} +2.94 & \cellcolor[rgb]{ .988,  .996,  1} +0.68 & \cellcolor[rgb]{ .914,  .957,  1} +4.12 & +2.72 \\
          & Python & \cellcolor[rgb]{ .788,  .894,  1} +10.08 & \cellcolor[rgb]{ .788,  .894,  1} +10.10 & \cellcolor[rgb]{ .906,  .953,  1} +4.53 & \cellcolor[rgb]{ .784,  .894,  1} +10.36 & \cellcolor[rgb]{ .792,  .898,  1} +9.86 & \cellcolor[rgb]{ .847,  .925,  1} +7.35 & \cellcolor[rgb]{ .902,  .953,  1} +4.68 & \cellcolor[rgb]{ .839,  .922,  1} +7.71 & \textbackslash{} & \cellcolor[rgb]{ 1,  .792,  .792} -5.71 & \cellcolor[rgb]{ .871,  .937,  1} +6.17 & \cellcolor[rgb]{ .945,  .973,  1} +2.77 & \cellcolor[rgb]{ .839,  .922,  1} +7.73 & +6.30 \\
          & Ruby  & \cellcolor[rgb]{ .851,  .925,  1} +7.06 & \cellcolor[rgb]{ .792,  .898,  1} +9.87 & \cellcolor[rgb]{ .914,  .957,  1} +4.20 & \cellcolor[rgb]{ .831,  .918,  1} +8.05 & \cellcolor[rgb]{ .804,  .902,  1} +9.44 & \cellcolor[rgb]{ .816,  .91,  1} +8.75 & \cellcolor[rgb]{ .984,  .992,  1} +0.86 & \cellcolor[rgb]{ .867,  .933,  1} +6.36 & \cellcolor[rgb]{ .992,  .996,  1} +0.48 & \textbackslash{} & \cellcolor[rgb]{ .835,  .918,  1} +7.97 & \cellcolor[rgb]{ .925,  .965,  1} +3.66 & \cellcolor[rgb]{ .788,  .894,  1} +10.19 & +6.41 \\
          & Scala & \cellcolor[rgb]{ .753,  .878,  1} +11.85 & \cellcolor[rgb]{ .753,  .878,  1} +11.68 & \cellcolor[rgb]{ .843,  .922,  1} +7.51 & \cellcolor[rgb]{ .765,  .882,  1} +11.15 & \cellcolor[rgb]{ .718,  .859,  1} +13.42 & \cellcolor[rgb]{ .996,  1,  1} +0.37 & \cellcolor[rgb]{ .898,  .949,  1} +4.87 & \cellcolor[rgb]{ .749,  .875,  1} +11.89 & \cellcolor[rgb]{ .863,  .933,  1} +6.63 & \cellcolor[rgb]{ .745,  .875,  1} +12.21 & \textbackslash{} & \cellcolor[rgb]{ .945,  .973,  1} +2.63 & \cellcolor[rgb]{ .804,  .902,  1} +9.36 & +8.63 \\
          & Swift & \cellcolor[rgb]{ .757,  .878,  1} +11.62 & \cellcolor[rgb]{ .725,  .863,  1} +12.98 & \cellcolor[rgb]{ .851,  .925,  1} +7.15 & \cellcolor[rgb]{ .718,  .859,  1} +13.46 & \cellcolor[rgb]{ .702,  .851,  1} +14.24 & \cellcolor[rgb]{ .847,  .925,  1} +7.28 & \cellcolor[rgb]{ .827,  .914,  1} +8.26 & \cellcolor[rgb]{ .729,  .867,  1} +12.92 & \cellcolor[rgb]{ .824,  .914,  1} +8.35 & \cellcolor[rgb]{ .776,  .89,  1} +10.73 & \cellcolor[rgb]{ .808,  .906,  1} +9.23 & \textbackslash{} & \cellcolor[rgb]{ .761,  .882,  1} +11.43 & +10.64 \\
          & TS    & \cellcolor[rgb]{ .749,  .875,  1} +11.88 & \cellcolor[rgb]{ .855,  .929,  1} +6.86 & \cellcolor[rgb]{ .847,  .925,  1} +7.26 & \cellcolor[rgb]{ .804,  .902,  1} +9.42 & \cellcolor[rgb]{ .69,  .847,  1} +14.80 & \cellcolor[rgb]{ .843,  .922,  1} +7.56 & \cellcolor[rgb]{ .91,  .957,  1} +4.28 & \cellcolor[rgb]{ .784,  .894,  1} +10.22 & \cellcolor[rgb]{ .867,  .933,  1} +6.44 & \cellcolor[rgb]{ .827,  .914,  1} +8.25 & \cellcolor[rgb]{ .945,  .973,  1} +2.62 & \cellcolor[rgb]{ .925,  .965,  1} +3.63 & \textbackslash{} & +7.77 \\
    \midrule
    \multicolumn{2}{c|}{\textbf{Mean}} & +10.46 & +11.31 & +7.46 & +11.18 & +13.20 & +9.41 & +6.58 & +12.44 & +6.78 & +9.38 & +6.51 & +4.82 & +11.44 & +9.30 \\
    \midrule
    \addlinespace
    \midrule
    \multicolumn{2}{c|}{\textbf{Python-specialized LLMs}} & \multicolumn{13}{c|}{\textbf{Target Programming Language}}                                            & \multirow{2}[2]{*}{\textbf{Mean}} \\
    \multicolumn{2}{c|}{\textbf{Pass@k}} & C++   & C\#   & Go    & Java  & JS    & Kotlin & Perl  & PHP   & Python & Ruby  & Scala & Swift & TS    &       \\
    \midrule
    \multicolumn{2}{c|}{\textbf{Baseline (without RACG)}} & 23.65  & 25.29  & 7.73  & 21.07  & 26.48  & 14.85  & 10.54  & 24.31  & 37.85  & 23.25  & 15.75  & 14.78  & 25.56  & 20.85 \\
    \midrule
    \multicolumn{1}{c}{\multirow{13}[2]{*}{\textbf{\makecell{Source\\Programming\\Language of\\Corpus}}}} & C++   & \textbackslash{} & \cellcolor[rgb]{ .878,  .941,  1} +7.19 & \cellcolor[rgb]{ .937,  .969,  1} +3.87 & +0.04 & \cellcolor[rgb]{ .89,  .945,  1} +6.40 & \cellcolor[rgb]{ 1,  .965,  .965} -0.84 & \cellcolor[rgb]{ 1,  .976,  .976} -0.51 & \cellcolor[rgb]{ .976,  .988,  1} +1.38 & \cellcolor[rgb]{ 1,  .973,  .973} -0.59 & \cellcolor[rgb]{ .957,  .98,  1} +2.54 & \cellcolor[rgb]{ .973,  .988,  1} +1.66 & +0.21 & \cellcolor[rgb]{ .961,  .98,  1} +2.49 & +1.99 \\
          & C\#   & \cellcolor[rgb]{ .851,  .925,  1} +8.72 & \textbackslash{} & \cellcolor[rgb]{ .949,  .976,  1} +3.04 & \cellcolor[rgb]{ .933,  .969,  1} +3.91 & \cellcolor[rgb]{ 1,  .922,  .922} -1.81 & \cellcolor[rgb]{ .957,  .98,  1} +2.66 & \cellcolor[rgb]{ 1,  .949,  .949} -1.20 & \cellcolor[rgb]{ 1,  .965,  .965} -0.78 & +0.04 & \cellcolor[rgb]{ .957,  .98,  1} +2.67 & \cellcolor[rgb]{ .973,  .988,  1} +1.76 & \cellcolor[rgb]{ 1,  .961,  .961} -0.91 & \cellcolor[rgb]{ .98,  .992,  1} +1.25 & +1.61 \\
          & Go    & \cellcolor[rgb]{ .973,  .988,  1} +1.69 & \cellcolor[rgb]{ 1,  .98,  .98} -0.38 & \textbackslash{} & \cellcolor[rgb]{ 1,  .937,  .937} -1.45 & \cellcolor[rgb]{ 1,  .702,  .702} -7.11 & \cellcolor[rgb]{ 1,  .816,  .816} -4.39 & \cellcolor[rgb]{ 1,  .89,  .89} -2.54 & \cellcolor[rgb]{ 1,  .784,  .784} -5.09 & \cellcolor[rgb]{ 1,  .855,  .855} -3.39 & \cellcolor[rgb]{ 1,  .718,  .718} -6.66 & \cellcolor[rgb]{ 1,  .843,  .843} -3.65 & \cellcolor[rgb]{ 1,  .706,  .706} -6.99 & \cellcolor[rgb]{ 1,  .824,  .824} -4.12 & -3.67 \\
          & Java  & \cellcolor[rgb]{ .878,  .941,  1} +7.08 & \cellcolor[rgb]{ .937,  .969,  1} +3.71 & \cellcolor[rgb]{ .918,  .961,  1} +4.97 & \textbackslash{} & +0.09 & \cellcolor[rgb]{ 1,  .918,  .918} -1.96 & \cellcolor[rgb]{ 1,  .961,  .961} -0.88 & \cellcolor[rgb]{ .988,  .996,  1} +0.69 & \cellcolor[rgb]{ 1,  .816,  .816} -4.36 & \cellcolor[rgb]{ .98,  .992,  1} +1.22 & \cellcolor[rgb]{ .976,  .988,  1} +1.52 & +0.21 & \cellcolor[rgb]{ .937,  .969,  1} +3.65 & +1.33 \\
          & JS    & \cellcolor[rgb]{ .875,  .937,  1} +7.37 & \cellcolor[rgb]{ .933,  .969,  1} +4.00 & \cellcolor[rgb]{ .957,  .98,  1} +2.71 & \cellcolor[rgb]{ 1,  .965,  .965} -0.79 & \textbackslash{} & \cellcolor[rgb]{ .976,  .988,  1} +1.45 & \cellcolor[rgb]{ 1,  .992,  .992} -0.18 & \cellcolor[rgb]{ .996,  1,  1} +0.35 & \cellcolor[rgb]{ 1,  .933,  .933} -1.52 & +0.14 & +0.14 & \cellcolor[rgb]{ 1,  .922,  .922} -1.87 & \cellcolor[rgb]{ .6,  .8,  1} +23.24 & +2.92 \\
          & Kotlin & \cellcolor[rgb]{ .973,  .988,  1} +1.78 & \cellcolor[rgb]{ .949,  .976,  1} +3.00 & \cellcolor[rgb]{ .949,  .976,  1} +3.09 & \cellcolor[rgb]{ .996,  1,  1} +0.44 & \cellcolor[rgb]{ .898,  .949,  1} +6.00 & \textbackslash{} & \cellcolor[rgb]{ 1,  .976,  .976} -0.55 & \cellcolor[rgb]{ 1,  .969,  .969} -0.69 & \cellcolor[rgb]{ 1,  .937,  .937} -1.48 & \cellcolor[rgb]{ .984,  .992,  1} +1.04 & \cellcolor[rgb]{ .973,  .988,  1} +1.61 & \cellcolor[rgb]{ 1,  .945,  .945} -1.23 & \cellcolor[rgb]{ .933,  .969,  1} +3.99 & +1.42 \\
          & Perl  & \cellcolor[rgb]{ .937,  .969,  1} +3.71 & \cellcolor[rgb]{ .996,  1,  1} +0.29 & \cellcolor[rgb]{ .996,  1,  1} +0.33 & \cellcolor[rgb]{ .976,  .988,  1} +1.45 & \cellcolor[rgb]{ 1,  .965,  .965} -0.79 & \cellcolor[rgb]{ .973,  .988,  1} +1.63 & \textbackslash{} & \cellcolor[rgb]{ 1,  .973,  .973} -0.60 & \cellcolor[rgb]{ .969,  .984,  1} +1.91 & \cellcolor[rgb]{ 1,  .941,  .941} -1.31 & \cellcolor[rgb]{ .988,  .996,  1} +0.76 & \cellcolor[rgb]{ 1,  .831,  .831} -3.95 & \cellcolor[rgb]{ .984,  .992,  1} +1.07 & +0.37 \\
          & PHP   & \cellcolor[rgb]{ .933,  .969,  1} +3.95 & \cellcolor[rgb]{ 1,  .988,  .988} -0.24 & \cellcolor[rgb]{ .988,  .996,  1} +0.88 & \cellcolor[rgb]{ 1,  .965,  .965} -0.79 & \cellcolor[rgb]{ 1,  .965,  .965} -0.79 & \cellcolor[rgb]{ .976,  .988,  1} +1.59 & \cellcolor[rgb]{ .988,  .996,  1} +0.88 & \textbackslash{} & \cellcolor[rgb]{ 1,  .91,  .91} -2.07 & \cellcolor[rgb]{ 1,  .914,  .914} -1.99 & +0.09 & \cellcolor[rgb]{ 1,  .898,  .898} -2.40 & \cellcolor[rgb]{ .992,  .996,  1} +0.47 & -0.04 \\
          & Python & \cellcolor[rgb]{ .965,  .984,  1} +2.26 & \cellcolor[rgb]{ 1,  .992,  .992} -0.14 & \cellcolor[rgb]{ .949,  .976,  1} +3.15 & \cellcolor[rgb]{ 1,  .933,  .933} -1.54 & \cellcolor[rgb]{ 1,  .894,  .894} -2.52 & \cellcolor[rgb]{ 1,  .965,  .965} -0.75 & +0.05 & \cellcolor[rgb]{ 1,  .878,  .878} -2.85 & \textbackslash{} & \cellcolor[rgb]{ 1,  .765,  .765} -5.58 & \cellcolor[rgb]{ 1,  .957,  .957} -0.95 & \cellcolor[rgb]{ 1,  .91,  .91} -2.08 & \cellcolor[rgb]{ .988,  .996,  1} +0.69 & -0.85 \\
          & Ruby  & \cellcolor[rgb]{ .937,  .969,  1} +3.76 & \cellcolor[rgb]{ .941,  .973,  1} +3.62 & \cellcolor[rgb]{ .988,  .996,  1} +0.77 & \cellcolor[rgb]{ 1,  .984,  .984} -0.35 & \cellcolor[rgb]{ .855,  .929,  1} +8.61 & \cellcolor[rgb]{ .976,  .988,  1} +1.49 & \cellcolor[rgb]{ .976,  .988,  1} +1.39 & \cellcolor[rgb]{ 1,  .894,  .894} -2.50 & \cellcolor[rgb]{ 1,  .965,  .965} -0.80 & \textbackslash{} & \cellcolor[rgb]{ 1,  .945,  .945} -1.23 & \cellcolor[rgb]{ 1,  .957,  .957} -0.96 & \cellcolor[rgb]{ .886,  .945,  1} +6.74 & +1.71 \\
          & Scala & \cellcolor[rgb]{ .984,  .992,  1} +1.11 & +0.14 & \cellcolor[rgb]{ .976,  .988,  1} +1.55 & \cellcolor[rgb]{ .976,  .988,  1} +1.54 & \cellcolor[rgb]{ 1,  .906,  .906} -2.16 & \cellcolor[rgb]{ 1,  .745,  .745} -6.07 & \cellcolor[rgb]{ 1,  .941,  .941} -1.34 & \cellcolor[rgb]{ 1,  .902,  .902} -2.29 & \cellcolor[rgb]{ 1,  .953,  .953} -1.10 & 0     & \textbackslash{} & \cellcolor[rgb]{ 1,  .949,  .949} -1.17 & \cellcolor[rgb]{ 1,  .918,  .918} -1.93 & -0.98 \\
          & Swift & \cellcolor[rgb]{ .996,  1,  1} +0.39 & \cellcolor[rgb]{ .965,  .984,  1} +2.10 & \cellcolor[rgb]{ .965,  .984,  1} +2.21 & \cellcolor[rgb]{ 1,  .878,  .878} -2.81 & \cellcolor[rgb]{ .988,  .996,  1} +0.75 & \cellcolor[rgb]{ 1,  .863,  .863} -3.27 & \cellcolor[rgb]{ 1,  .965,  .965} -0.79 & \cellcolor[rgb]{ 1,  .973,  .973} -0.60 & \cellcolor[rgb]{ 1,  .957,  .957} -0.97 & \cellcolor[rgb]{ 1,  .933,  .933} -1.59 & \cellcolor[rgb]{ 1,  .824,  .824} -4.13 & \textbackslash{} & \cellcolor[rgb]{ .965,  .984,  1} +2.10 & -0.55 \\
          & TS    & \cellcolor[rgb]{ .918,  .961,  1} +4.96 & \cellcolor[rgb]{ 1,  .937,  .937} -1.43 & \cellcolor[rgb]{ .918,  .961,  1} +4.86 & \cellcolor[rgb]{ .988,  .996,  1} +0.83 & \cellcolor[rgb]{ .733,  .867,  1} +15.53 & \cellcolor[rgb]{ 1,  .929,  .929} -1.63 & \cellcolor[rgb]{ 1,  .89,  .89} -2.54 & \cellcolor[rgb]{ 1,  .82,  .82} -4.23 & \cellcolor[rgb]{ 1,  .914,  .914} -2.03 & \cellcolor[rgb]{ 1,  .953,  .953} -1.09 & \cellcolor[rgb]{ 1,  .937,  .937} -1.47 & \cellcolor[rgb]{ 1,  .812,  .812} -4.48 & \textbackslash{} & +0.61 \\
    \midrule
    \multicolumn{2}{c|}{\textbf{Mean}} & +3.90 & +1.82 & +2.62 & +0.04 & +1.85 & -0.84 & -0.69 & -1.44 & -1.37 & -0.88 & -0.32 & -2.13 & +3.30 & +0.45 \\
    \bottomrule
    \end{tabular}%
    }
    \caption{RACG performance compared to baseline (without RACG) when the retrieval corpus and generation task involve different programming languages on \textit{Large Multilingual Code Dataset} in \textit{Doc w/o NL} setting.}
    \label{tab:pass@k_cross_enhance_real_big}
\end{table*}

\subsection{Dataset and Setup}

Code corpora on the Internet most exist as isolated fragments without corresponding comments or aligned NL descriptions~\cite{surveycodecomments}. For simulation of this setting (\ie \textit{Doc w/o NL} in Figure~\ref{fig:Experimental_setting}), we provide a \textit{pure code corpus} version of the \textit{Large Multilingual Code Dataset} dataset stripped of NL comments. We also annotate 1 golden document per \textit{PL} for each query. This design ensures that each query corresponds to 13 golden documents across \textit{PL}s, enabling calculation of retrieval precision and recall rates. The large dataset scale and intentional information deprivation create challenging yet realistic conditions for examining cross-lingual knowledge transfer mechanisms.

To explore the different retrieval performance in this \textit{Doc w/o NL} setting, we investigate three distinct retrieval paradigms in RACG:
\textit{First}, lexical sparse retrieval (\eg, BM25), which is computationally efficient and relies on keyword matching and term frequency statistics. 
\textit{Second}, generic text embedding models, where code is treated as normal text using a pre-trained language model (\eg, BERT-style architectures) to encode. 
\textit{Third}, domain-specific code retrievers, which are trained on NL-to-code alignment tasks and can bridge NL intent and code semantics.

Specifically, we evaluate retrieval strategies on the \textit{Large Multilingual Code Dataset} in \textit{Doc w/o NL} setting, where, for each query, one most relevant code snippet is annotated per \textit{PL}.
The evaluated models span the three paradigms: \textit{BM25} method for sparse retrieval; \textit{Sup-SimCSE-BERT-large-uncased}~\cite{gao2021simcse}, \textit{BGE-large-en-v1.5}~\cite{bge_embedding} and the larger \textit{Qwen3-Embedding-8B}~\cite{qwen3embedding} models for generic text embedding; and \textit{CodeRankEmbed}~\cite{suresh2024cornstack} model for domain-specific code retrieval. We use Precision@K and Recall@K as metrics to assess the code retrieval effectiveness.

\begin{table*}[h]
  \centering
  \resizebox{\textwidth}{!}{
    \begin{tabular}{l|l|cccc}
    \toprule
    \textbf{Strategies} & \textbf{Retrieval Model} & \textbf{Precison@5} & \textbf{Precison@10} & \textbf{Recall@10} & \textbf{Recall@20} \\
    \midrule
    Lexical Sparse Retrieval & BM25  & 6.57\% & 4.87\% & 3.74\% & 4.79\% \\
    \midrule
    \multirow{3}[2]{*}{Generic Text Embedding} & Sup-SimCSE-BERT-large & 18.65\% & 13.75\% & 10.58\% & 14.44\% \\
          & BGE-large-en-v1.5 & 56.18\% & 44.64\% & 34.34\% & 46.05\% \\
          & Qwen3-Embedding-8B & 77.29\% & 66.99\% & 51.53\% & 68.38\% \\
    \midrule
    Domain-specific NL-to-Code Alignment & CodeRankEmbed & 91.60\% & 87.72\% & 67.48\% & 88.04\% \\
    \bottomrule
    \end{tabular}%
    }
    \caption{Effectiveness of different retrieval strategies in RACG for the \textit{Doc w/o NL} setting.}
    \label{tab:retrieval}
\end{table*}%

\subsection{Natural Language Plays a Minor Role under Strong Code Retrieval.}

Table~\ref{tab:pass@k_cross_enhance_real_big} shows RACG performance compared to baseline (without RACG) when the retrieval corpus and generation task involve different \textit{PL}s on \textit{Large Multilingual Code Dataset} in \textit{Doc w/o NL} setting, while effectiveness of different retrieval strategies in RACG for the \textit{Doc w/o NL} setting is reported in Table~\ref{tab:retrieval}. 
In RACG, removing NL from the retrieval corpus has little impact on cross-lingual knowledge transfer ablility. In this case, domain-specific retriever exhibits best retrieval performance.

\begin{mdframed}[style=MyFrame]
\textbf{Finding 7:} NL in the code corpus (\eg~code comments) has little impact on knowledge transfer performance cross \textit{PL}s in RACG.
\end{mdframed}

Our investigation reveals a counter-intuitive finding regarding the role of NL in RACG. While it is often assumed that NL documentation (e.g., comments, docstrings) serves as a critical semantic bridge between differing \textit{PL}s, our empirical data suggests its impact is secondary to the code itself.

Table~\ref{tab:pass@k_cross_enhance_real_big} and Table~\ref{tab:pass@k_cross_enhance_ideal_big} share the same experimental setup, except that Table~\ref{tab:pass@k_cross_enhance_real_big} uses the \textit{Doc w/o NL} setting. By comparing the two tables, it can be seen that after removing NL, the knowledge contained in the corpus decreases. The overall knowledge transfer effect of RACG declines, but is very slight (+11.67\%→+9.30\% for multi-lingual LLMs overall performance and +1.04\%→+0.45\% for Python-specialized LLMs).

For multi-lingual LLMs, performance gap is merely 2.37\%. While the presence of comments does provide contextual cues that aid retrieval and generation, the vast majority of the improvement (over 9\%) is driven solely by the semantic information encoded within the programming logic itself. The "heat" of the improvement (represented by the blue cells in the Tables~\ref{tab:pass@k_cross_enhance_real_big}) remains intense even when NL is stripped away.

\begin{mdframed}[style=MyFrame]
\textbf{Finding 8:} Different retrieval strategies yield varied effectiveness on pure code corpus. Domain-specific retrievers perform the best.
\end{mdframed}

As shown in Table~\ref{tab:retrieval}, significant performance variations emerge across strategies. The sparse retriever \textit{BM25}, based on bag-of-words matching, performs worst with only 6.57\%Precision@5 and 4.79\%Recall@20, confirming the inadequacy of pure lexical matching for code. The generic embedding models show a clear performance hierarchy correlating with their general text embedding strength. \textit{Sup-SimCSE-BERT-large-uncased}, a contrastively fine-tuned BERT model, achieves moderate gains (18.65\%P@5, 14.44\%R@20). The more advanced general-purpose embedders perform better: \textit{BGE-large-en-v1.5} reaches 56.18\%P@5 and 46.05\%R@20, while the larger model \textit{Qwen3-Embedding-8B} further improves to 77.29\%P@5 and 68.38\%R@20, indicating that powerful general text models can capture some code semantics. In contrast, the domain-specific \textit{CodeRankEmbed} delivers the best results, attaining approximately 90\%in both precision and recall metrics (91.60\%P@5, 88.04\%R@20), significantly outperforming all general-purpose models.

\section{Related Work}

\subsection{Retrieval-Augmented Code Generation}

Recent years have witnessed significant advances in RACG, where external contexts and documentation are leveraged to enhance code tasks. Classic works such as REDCODER~\cite{parvez2021retrieval}, ReACC~\cite{lu2022reacc}, and DocPrompt~\cite{zhou2023docprompting} demonstrate its efficacy for code generation, summarization and completion.
Subsequent research expand RACG methodologies and benchmarks, yielding notable contributions~\cite{li2023large, li2025building, dutta2024rar, tan2024prompt, su2024evor}. However, existing works such as~\cite{wang2024coderagbench, du2024codegrag, gao2024preference} remain limited, only focusing on 1-2 mainstream \textit{PL}s.
For instance, CodeRAG-Bench~\cite{wang2024coderagbench} establishes a comprehensive evaluation framework for Python code  RACG; CodeGRAG~\cite{du2024codegrag} explore Python and C++ RACG which use graphical view of code blocks; RRG~\cite{gao2024preference} introduces a code refactorer module in Python and Java RACG.
Our work goes beyond existing works by investigating cross-lingual RACG across 13 \textit{PL}s with various settings to explore knowledge transfer.

\subsection{Multi-\textit{PL} Code Generation Benchmark}

The emergence of multi-lingual evaluation benchmarks has driven progress in cross-lingual code generation research. HumanEval-X~\cite{zheng2023codegeex}, MultiPL-E~\cite{cassano2022multipl} and CruxEval-X~\cite{xu2025cruxeval} enable parallel evaluation across multiple \textit{PL}s by translating existing problems, and McEval~\cite{chai2024mceval} further enhances the diversity of evaluation data. 
However, existing studies predominantly focus on benchmarking performance, leaving cross-lingual knowledge transfer mechanisms underexplored.
The work~\cite{athiwaratkun2022multilingual} explains the mutual augmentation capability between different \textit{PL} corpora through the zero-shot code translation ability of LLMs, which is entirely distinct from the perspective of our study. In comparison, we focuse on cross-lingual knowledge transfer challenges in complete RACG.
\section{Conclusion}

In this paper, we investigated knowledge transfer across \textit{PL}s through the large-scale empirical study. By constructing a high-quality dataset spanning 13 \textit{PL}s of nearly 14k code generation instances, we explore three RQs about knowledge transfer in RACG across \textit{PL}s. This study reveals the complexity and imbalance of cross-lingual knowledge transfer in RACG, providing a roadmap for effective multi-lingual code generation assistant.

\section*{Limitations}

There are two primary limitations in our study. First, the LLMs evaluated in the RACG are all of relatively small scale (with at most 14B parameters), which may not be fully representative of all code-related models. Second, the scale of the current dataset is constrained, comprising approximately 1,000 samples per programming language. This limitation may affect the comprehensiveness and statistical robustness of the evaluation across diverse linguistic features.
Future work can improve the study by using larger code LLMs and benchmark expended through large-scale synthetic data generation techniques.

\section*{Acknowledgements}

We sincerely thank the reviewers for their insightful comments and valuable suggestions. This work was supported by the National Key R\&D Program of China (2024YFC3308000), the Natural Science Foundation of China (No. 62506354, 62306303, 62476265).

\bibliography{bibref}

\appendix

\begin{table*}[htb]
  \centering
  \resizebox{0.95\textwidth}{!}{
    \begin{tabular}{cc|ccc|c}
    \toprule
    \multicolumn{2}{c|}{\multirow{2}[2]{*}{\textbf{Pass@k}}} & \multicolumn{3}{c|}{\textbf{Multi-lingual LLMs}} & \multirow{2}[2]{*}{\textbf{Mean}} \\
    \multicolumn{2}{c|}{} & CodeLlama-7B & Deepseek-Coder-6.7B & Qwen2.5-Coder-7B &  \\
    \midrule
    \multicolumn{1}{c}{\multirow{13}[2]{*}{\textbf{\makecell{Source\\Programming\\Language of\\Corpus}}}} & C++   & +15.62 & +11.56 & +8.70 & +11.96 \\
          & C\#   & +16.38 & +12.59 & +8.98 & +12.65 \\
          & Go    & +13.87 & +11.78 & +9.24 & +11.63 \\
          & Java  & +17.18 & +12.96 & +9.92 & +13.35 \\
          & JS    & +10.66 & +8.25 & +6.73 & +8.55 \\
          & Kotlin & +11.34 & +8.79 & +3.71 & +7.95 \\
          & Perl  & +14.57 & +13.05 & +9.57 & +12.40 \\
          & PHP   & +4.45 & +3.36 & +0.36 & +2.72 \\
          & Python & +5.06 & +8.18 & +5.67 & +6.30 \\
          & Ruby  & +6.05 & +8.05 & +5.13 & +6.41 \\
          & Scala & +9.38 & +10.67 & +5.84 & +8.63 \\
          & Swift & +12.48 & +12.34 & +7.09 & +10.64 \\
          & TS    & +10.38 & +7.34 & +5.58 & +7.77 \\
    \midrule
    \multicolumn{2}{c|}{\textbf{Mean}} & +11.34 & +9.92 & +6.66 & +9.30 \\
    \bottomrule
    \end{tabular}%
    }
    \caption{Multi-lingual LLMs RACG performance compared to baseline (without RAG) when the retrieval corpus and generation task involve different programming languages on \textit{Large Multilingual Code Dataset} in \textit{Doc w/o NL} setting.}
  \label{tab:pass@k_cross_enhance_real_big_bias}%
\end{table*}%

\section{Prompt Template for Benchmark Evaluation}
\label{sec:prompt}

\begin{figure}[h!]
\centering
\includegraphics[width=0.95\columnwidth]{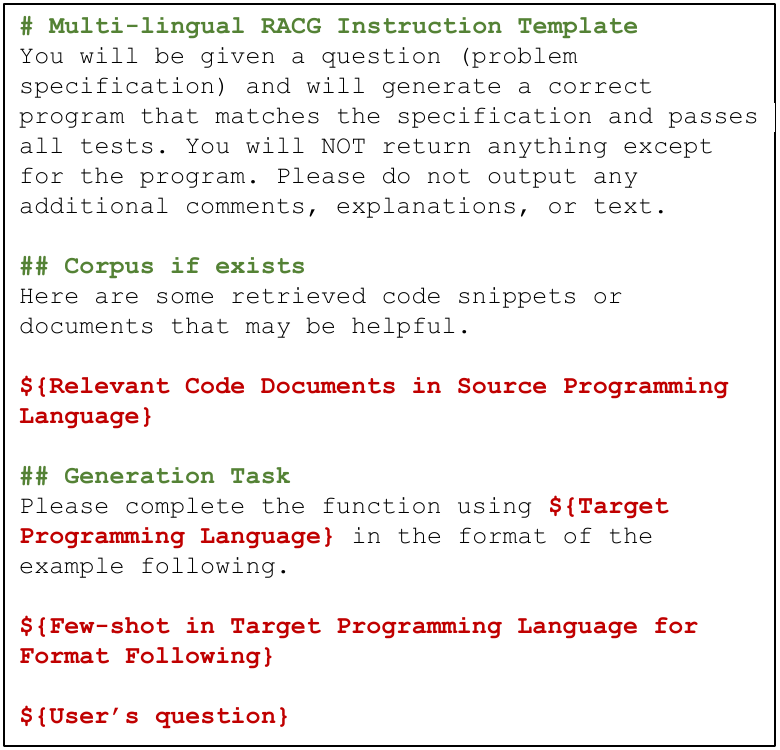}
\caption{Unified prompt template for LLMs to generate code in multi-lingual RACG.}
\label{fig:RAG_prompt_for_LLM}
\end{figure}

We evaluate the LLMs by using a unified prompt template (as illustrated in Figure~\ref{fig:RAG_prompt_for_LLM}) that instructs them to utilize the code corpus, thereby enhancing code generation.

\section{Statistic of \textit{Large Multilingual Code Dataset}}
\label{sec:large_dataset}

To conduct large-scale empirical experiments and explore the knowledge transfer effects between more \textit{PL} pairs, we construct the \textit{Large Multilingual Code Dataset} based on existing datasets, which will be used as both an evaluation benchmark and a retrieval library. The dataset has nearly 14k high-quality code generation instances across 13 \textit{PL}s. As shown in Table~\ref{tab:large_dataset}, each \textit{PL} has approximately 1,000 datapoints.

\begin{table}[h!]
\centering
\resizebox{0.95\columnwidth}{!}{
\begin{tabular}{cc}
\toprule
\textbf{Programming Language} & \textbf{Correct Instance} \\
\midrule
Python & 1181 \\
Kotlin & 1071 \\
Java & 1139 \\
Ruby & 1103 \\
JavaScript (JS) & 1133 \\
PHP & 1158 \\
TypeScript (TS) & 1059 \\
C++ & 1038 \\
C\# & 1050 \\
Go & 905 \\
Perl & 1082 \\
Scala & 1054 \\
Swift & 937 \\
\midrule
Total & 13910 \\
\bottomrule
\end{tabular}
}
\caption{Instance distribution across 13 programming languages in \textit{Large Multilingual Code Dataset} for RACG (Total 13,910 instances)}
\label{tab:large_dataset}
\end{table}

\section{Dataset Bias for Data-Generating LLM}
\label{sec:bias}

One might be concerned that the reference solutions generated by \textit{Qwen-72B} could favor models from the \textit{Qwen} series. For example, Table~\ref{tab:pass@k_cross_enhance_real_big} reports the average results for multilingual LLMs in the \textit{Doc w/o NL} setting, where the retrieval corpus and generation task involve different programming languages on the \textit{Large Multilingual Code Dataset}. To address this concern more thoroughly, we present the per-model Pass@K improvements for three models in Table~\ref{tab:pass@k_cross_enhance_real_big_bias}. The results show that \textit{Qwen2.5-Coder-7B} obtains a smaller gain (6.66\%↑) compared to \textit{CodeLlama-7B} (11.34\%↑) and \textit{DeepSeek-Coder-6.7B} (9.92\%↑). Since the Qwen-Coder model does not receive a larger advantage, this indicates that the concern about bias is unwarranted. This conclusion is also supported by prior work~\cite{gu2024cruxeval,xu2025cruxeval}, which suggests that such concerns are unnecessary.

\section{Evaluation Metrics}
\label{sec:pass_at_5}

We employ greedy decoding (Temperature=0.0) to ensure the reproducibility of our experimental results. In deterministic (greedy) settings, Pass@1 provides a clear and reliable measurement of a model’s ability to follow retrieved instructions and generate correct code in a single attempt, which is crucial for practical RACG applications. While security and efficiency are important, this study primarily focuses on the feasibility and effectiveness of cross-lingual knowledge transfer. Correctness (Pass@K) serves as the primary indicator of successful knowledge transfer across programming language silos. We appreciate the reviewer’s suggestion, and we will continue exploring code security and style in future work.

Refering the settings from previous work~\cite{xu2025cruxeval, zhuobigcodebench}, we evaluated Pass@1 (Temperature = 0.0, SampleN = 1) and Pass@5 (Temperature = 0.8, SampleN = 5). We conduct experiments using a subset of the dataset on Multi-lingual LLMs (\textasciitilde 7B). The results, briefly summarized in the Table~\ref{tab:pass_at_5}, show that increasing the number of samples improves the LLM's performance. In this case, RACG remains a highly effective method for performance enhancement and continues to exhibit consistent patterns in the Pass@K (K=1,5) results. These findings further support the conclusions of our study.

\begin{table}[htbp]
  \centering
  \resizebox{0.95\columnwidth}{!}{
    \begin{tabular}{cc|cc}
    \toprule
    \multicolumn{2}{c|}{\textbf{Multi-lingual LLMs (\textasciitilde 7B)}} & \multirow{2}[2]{*}{\textbf{Pass@1}} & \multirow{2}[2]{*}{\textbf{Pass@5}} \\
    \multicolumn{2}{c|}{\textbf{Window Size K=3}} &       &  \\
    \midrule
    \multicolumn{2}{c|}{\textbf{Baseline (without RAG)}} & 55.75 & 70.42 \\
    \midrule
    \multicolumn{1}{c}{\multirow{8}[2]{*}{\textbf{\makecell{Source\\Programming\\Language of\\Corpus}}}} & Python & +13.53 & +12.38 \\
          & Java  & +16.86 & +15.90 \\
          & JS    & +13.52 & +14.48 \\
          & TS    & +13.05 & +11.90 \\
          & Ruby  & +12.76 & +12.48 \\
          & C++   & +19.71 & +18.57 \\
          & Go    & +17.71 & +13.33 \\
          & Swift & +16.48 & +12.47 \\
    \midrule
    \multicolumn{2}{c|}{\textbf{Mean}} & +15.45 & +13.94 \\
    \bottomrule
    \end{tabular}%
  }
  \caption{RACG performance compared to baseline (without RACG) when the retrieval corpus and generation task involve different programming languages on the subset of \textit{Large Multilingual Code Dataset} in \textit{Doc} setting.}
  \label{tab:pass_at_5}%
\end{table}%

\section{Parameter Configuration}
\label{sec:window_size}

The retrieval window size is fixed at K=3 to strictly control variables across all evaluated models. This specific value is chosen as a pragmatic trade-off to accommodate the context length limitations of certain models, while still providing sufficient external knowledge for effective RACG.

We also conduct experiments on a subset of the dataset to explore the impact of the context retrieval window size K (K=1, 3, 5). The results, briefly summarized in the Table~\ref{tab:window_size}, show that when the context length allows, introducing more external knowledge can help the model generate better code. These findings further support the conclusions of our study.

\begin{table}[htbp]
  \centering
  \resizebox{0.95\columnwidth}{!}{
    \begin{tabular}{cc|ccc}
    \toprule
    \multicolumn{2}{c|}{\textbf{Multi-lingual LLMs (\textasciitilde 7B)}} & \multirow{2}[2]{*}{\textbf{K=1}} & \multirow{2}[2]{*}{\textbf{K=3}} & \multirow{2}[2]{*}{\textbf{K=5}} \\
    \multicolumn{2}{c|}{\textbf{Pass@1}} &       &       &  \\
    \midrule
    \multicolumn{2}{c|}{\textbf{Baseline (without RAG)}} & \multicolumn{3}{c}{55.75} \\
    \midrule
    \multicolumn{1}{c}{\multirow{8}[2]{*}{\textbf{\makecell{Source\\Programming\\Language of\\Corpus}}}} & Python & +13.53 & +13.71 & +15.52 \\
          & Java  & +16.86 & +17.14 & +16.19 \\
          & JS    & +13.52 & +15.33 & +16.19 \\
          & TS    & +13.05 & +13.91 & +14.57 \\
          & Ruby  & +12.76 & +14.95 & +15.24 \\
          & C++   & +19.71 & +20.76 & +21.33 \\
          & Go    & +17.71 & +18.67 & +18.95 \\
          & Swift & +16.48 & +16.76 & +16.38 \\
    \midrule
    \multicolumn{2}{c|}{\textbf{Mean}} & +15.45 & +16.41 & +16.80 \\
    \bottomrule
    \end{tabular}%
  }
  \caption{RACG performance compared to baseline (without RACG) when the retrieval corpus and generation task involve different programming languages on the subset of \textit{Large Multilingual Code Dataset} in \textit{Doc} setting.}
  \label{tab:window_size}%
\end{table}%

\section{Model Size}
\label{sec:model_size}

\begin{table}[htbp]
  \centering
  \resizebox{0.95\columnwidth}{!}{
    \begin{tabular}{cc|ccc}
    \toprule
    \multicolumn{2}{c|}{\textbf{Qwen2.5-Coder-14B-Instruct}} & \multirow{2}[2]{*}{\textbf{K=1}} & \multirow{2}[2]{*}{\textbf{K=3}} & \multirow{2}[2]{*}{\textbf{K=5}} \\
    \multicolumn{2}{c|}{\textbf{Pass@1}} &       &       &  \\
    \midrule
    \multicolumn{2}{c|}{\textbf{Baseline (without RAG)}} & \multicolumn{3}{c}{72.25} \\
    \midrule
    \multicolumn{1}{c}{\multirow{8}[2]{*}{\textbf{\makecell{Source\\Programming\\Language of\\Corpus}}}}
          & Python & +8.86 & +10.57 & +10.29 \\
          & Java  & +8.00 & +9.43 & +7.43 \\
          & JS    & +8.00 & +12.00 & +10.29 \\
          & TS    & +8.86 & +9.14 & +9.14 \\
          & Ruby  & +6.57 & +9.71 & +9.43 \\
          & C++   & +11.14 & +12.57 & +12.86 \\
          & Go    & +10.00 & +12.00 & +12.29 \\
          & Swift & +11.43 & +10.29 & +10.29 \\
    \midrule
    \multicolumn{2}{c|}{\textbf{Mean}} & +9.11 & +10.71 & +10.25 \\
    \bottomrule
    \end{tabular}%
  }
  \caption{RACG performance compared to baseline (without RACG) when the retrieval corpus and generation task involve different programming languages on the subset of \textit{Large Multilingual Code Dataset} in \textit{Doc} setting.}
  \label{tab:model_size}%
\end{table}%

With similar model sizes controlled for variable control, we evaluate a representative set of code LLMs around 7B size. Given the reviewers' interest in seeing experimental results with larger code LLMs, we add experiments with a larger code LLM (Qwen2.5-Coder-14B-Instruct) using a subset of the datasets. We also explore the impact of different retrieval window sizes (K=1, 3, 5). The results, briefly summarized in the Table~\ref{tab:model_size}, show that larger models exhibit stronger code generation capabilities and still achieve significant performance gains through RACG. These findings continue to support the conclusions of our study.

\end{document}